\documentclass[aps,nofootinbib,showpacs,preprintnumbers,amsmath,amssymb]{revtex4}
\def\be{\begin{equation}}
\def\ee{\end{equation}}
\def\ba{\begin{eqnarray}}
\def\ea{\end{eqnarray}}
\newcommand{\A}{{\mathcal{A}}}
\newcommand{\tA}{{\widetilde {\mathcal{A}}}}
\newcommand{\ta}{{\widetilde a}}
\newcommand{\tlA}{{\mathfrak A}}
\usepackage{graphics}
\usepackage{graphicx}
\usepackage{dcolumn}
\usepackage{bm}
\usepackage{epsfig}
\usepackage{graphicx}
\usepackage{multirow}
\usepackage{dcolumn}
\usepackage{graphicx,epsfig}%

\begin{document}

\preprint{arXiv:1203.6897v2,preprint USM-TH-298, Phys.~Rev.~D85,114043 (2012)}

\title{Extended analytic QCD model with perturbative QCD behavior 
at high momenta\footnote{
Dedicated to Olivier Espinosa (1961-2010).}}
\author{C\'esar Ayala$^1$}
\author{Carlos Contreras$^1$}
\author{Gorazd Cveti\v{c}$^1$$^2$}

\affiliation{$^1$Department of Physics, Universidad T{\'e}cnica Federico
Santa Mar{\'\i}a (UTFSM), Valpara{\'\i}so, Chile\\
$^2$Centro Cient\'{\i}fico-Tecnol\'ogico de Valpara\'{\i}so, UTFSM, Chile}

\date{\today}

\begin{abstract}

In contrast to perturbative QCD, the analytic QCD models have
running coupling whose analytic properties correctly mirror 
those of spacelike observables.
The discontinuity (spectral) function of such running coupling is
expected to agree with the perturbative case at large timelike
momenta; however, at low timelike momenta it is not known.
In the latter regime, we parametrize the unknown behavior of 
the spectral function as a sum of (two) delta functions; while the 
onset of the perturbative behavior of the spectral function 
is set to be $1.0$-$1.5$ GeV. This 
is in close analogy with the ``minimal hadronic ansatz'' used
in the literature for modeling spectral functions of correlators.
For the running coupling itself, we impose the condition
that it basically merges with the perturbative coupling at high
spacelike momenta. In addition, we require that the well-measured
nonstrange semihadronic $(V+A)$ tau decay ratio value be reproduced by 
the model. We thus obtain a QCD framework which is basically 
indistinguishable from perturbative QCD at high momenta ($Q > 1$ GeV), 
and at low momenta it respects the basic analyticity properties of 
spacelike observables as dictated by the general principles of the
local quantum field theories.

\end{abstract}

\maketitle

\section{Introduction}
\label{sec:intr}

The general principles of the local quantum field theories imply 
\cite{BS,Oehme} that the spacelike observables ${\cal D}(Q^2)$ are
analytic functions of $Q^2$ in the complex $Q^2$-plane,
with the nonanalyticity allowed only on the timelike semiaxis
$Q^2 < 0$. Here, $-q^2\equiv Q^2$ where $q$ is the typical momentum of
the considered proccess, e.g., momentum of the exchanged photon
in deep inelastic scattering. However, in perturbative QCD (pQCD), 
the running coupling $a_{\rm pt}(Q^2) \equiv \alpha_s(Q^2)/\pi$ in most
renormalization schemes has (Landau) singularities
in the complex $Q^2$-plane close to the origin
($|Q^2| \alt 1 \ {\rm GeV}^2$) which do not reflect
the aforementioned analytic properties. This fact represents
a serious, and often underestimated, conceptual and practical problem.
The (spacelike) observables evaluated in pQCD, as
truncated power series of $a_{\rm pt}(\kappa Q^2)$ (with $\kappa \sim 1$), 
do not possess the aformentioned analyticity properties; 
furthermore, for low values of $|Q^2|$ ($\sim 1 \ {\rm GeV}^2$),
they are numerically unreliable due to the vicinity of the scale
$\kappa Q^2$ to the unphysical (Landau) singularities. This
numerical unreliability reflects itself in a strong dependence
on the renormalization scale ($\leftrightarrow \kappa$) and scheme.

On the other hand, studies using Dyson-Schwinger equations 
\cite{DSE1,DSE2} and lattice calculations \cite{latt} indicate
that the QCD running coupling is finite (``conformal'') at $Q^2=0$,
and has, at least at positive $Q^2$ values, no Landau singularities. 

The problem of Landau poles in the QCD coupling
was first addressed in a systematic manner about 15 years 
ago by the authors of \cite{ShS,MSS,Sh}, who constructed
and used an analytic QCD coupling parameter $\A_1^{\rm (MA)}(Q^2)$
closely based on the perturbative coupling parameter $a_{\rm pt}(Q^2)$: 
in the dispersive integral expression for $a_{\rm pt}(Q^2)$, they
removed the integration over the offending spacelike 
discontinuity cut (i.e., at $-Q^2 = \sigma < 0$),
while keeping the discontinuity (spectral) function
$\rho_1^{\rm (pt)}(\sigma) \equiv {\rm Im} \ a_{\rm pt}(Q^2 = -\sigma - i \epsilon)$
unchanged on the timelike discontinuity cut, i.e., at $\sigma \geq 0$.
Therefore, we can call this the Minimal Analytic (MA) model. 
The authors of \cite{ShS,MSS,Sh} called it the
Analytic Perturbation Approach (APT), and provided an
analogous method of construction of analytic analogs 
$\A_n^{\rm (MA)}(Q^2)$ of the powers $a_{\rm pt}(Q^2)^n$ for $n=2,3,\ldots$.
A method of construction of analytic analogs $\A_{\nu}^{\rm (MA)}(Q^2)$
of noninteger powers $a_{\rm pt}(Q^2)^{\nu}$ for MA (APT) was developed
by the authors of \cite{BMS}. 
Yet another analytic QCD model, based on the minimal analytization 
of the beta function $d a_{\rm pt}(Q^2)/d \ln Q^2$, was constructed 
and used in \cite{Nesterenko}.

On the other hand, a more general approach of constructing 
analytic QCD coupling $\A_1(Q^2)$, based on a given choice
of the discontinuity function $\rho_1(\sigma) \equiv 
{\rm Im} \ \A_1(Q^2=-\sigma - i \epsilon)$ (for $\sigma > 0$) was
emphasized in \cite{CV1,CV2} and (in a more
specific context) in \cite{NestPapa}.
The spacelike coupling $\A_1(Q^2)$ is then constructed 
from $\rho_1(\sigma)$ by the usual dispersion relation
\be
\A_1(Q^2) 
= \frac{1}{\pi} \int_{0}^{+\infty} \ d \sigma 
\frac{ \rho_1(\sigma) }{(\sigma + Q^2)} \ .
\label{dispA1}
\ee
From our standpoint it is this analytic coupling $\A_1$ 
(or equivalently: the spectral function $\rho_1$) 
that defines the analytic QCD (anQCD) model.
The construction of analytic analogs $\A_n(Q^2)$ of integer powers 
$a_{\rm pt}(Q^2)^n$, applicable to any such analytic QCD model,
was performed in \cite{CV1,CV2}, using the relations
between the logarithmic derivatives $\ta_{{\rm pt},n}(Q^2)$
($\propto d^{n-1} a_{\rm pt}(Q^2)/d (\ln Q^2)^{n-1}$)
 and powers
$a_{\rm pt}(Q^2)^k$
\be
a_{\rm pt}^n = \ta_{{\rm pt}, n} + 
\sum_{m \geq 1} {\widetilde k}_m(n) \ta_{{\rm pt},n+m} \ ,
\label{antan}
\ee
and\footnote{
The recurrence relations leading to the above relations,
within the context of the MA (APT) model of  \cite{ShS,MSS,Sh}, 
were given in \cite{Shirkov:2006nc,Shirkov}.} 
imposing the condition of analogy on the
logarithmic derivatives of $a_{\rm pt}$ and of $\A_1$: 
$\ta_{{\rm pt},n+m}  \mapsto \tA_{n+m}$.
This condition was shown to be imperative, in
order to keep the renormalization scale and scheme
dependence of the resulting truncated analytic series
(in terms of $\tA_{n}$, or $\A_n$) of physical observables under control.
The construction of the power analogs $\A_n$ of $a_{\rm pt}^n$
in general analytic QCD models was obtained thus from $\A_1$ via 
the relations analogous to (\ref{antan})\footnote{
The construction of higher power analogs $\A_n$
as linear operations on $\A_1$ (not as: $\A_1^n$)
incorporates a nice functional property: its compatibility 
with linear integral transformations, such as Fourier or Laplace 
\cite{Shirkov:1999np}.}
\be
\A_n = \tA_n + \sum_{m \geq 1} {\widetilde k}_m(n) \tA_{n+m} \ .  
 \label{AntAn}
\ee
The extension of this construction to noninteger power analogs
$\A_{\nu}(Q^2)$, for general analytic QCD models, was performed in
\cite{GCAK}.

So, from our standpoint, it remains an outstanding problem to
obtain or construct the most acceptable analytic coupling
$\A_1(Q^2)$, or equivalently, the spectral function
$\rho_1(\sigma) = {\rm Im} \ \A_1(Q^2=-\sigma - i \epsilon)$ (for $\sigma > 0$).
It is reasonable to assume that at large 
$\sigma$ ($ > 1 \ {\rm GeV}^2$) we have
$\rho_1(\sigma) = \rho_1^{\rm (pt)}(\sigma)$, i.e., the spectral
function agrees with the pQCD result. On the other hand,
at low $\sigma \alt 1 \ {\rm GeV}^2$, the exact behavior of $\rho_1(\sigma)$
is unknown.

The construction of $\A_1(Q^2)$ can be performed in two different
ways. One way is to construct first the beta function $\beta(\A_1)
= d \A_1/d \ln Q^2$ as function of $\A_1$. This approach is
convenient if we take the position that $\beta(x)$
is an analytic function of $x$ at $x=0$. In such a case,
it tuns out that we obtain perturbative QCD, i.e., $\A_n = \A_1^n$. 
However, in this case,
after ensuring additionally the analyticity of $\A_1(Q^2)$ 
as a function of $Q^2$, it turns out to be very difficult
to reproduce the correct measured value of the tau lepton
(nonstrange) semihadronic $(V+A)$ decay ratio $r_{\tau} \approx 0.20$,
cf.~\cite{CKV}. In fact, as shown in \cite{CKV},
the large enough value $r_{\tau}$ can be obtained
in the perturbative analytic QCD frameworks
only at an (unacceptable?) price of choosing a renormalization
scheme with increasingly large $\beta_j$ coefficients, which makes
the analytic perturbation series of observables convergent only
when up to four terms are included, and the fifth (${\rm N}^4{\rm LO}$)
term in the expansion shows an explosive increase.

Another way is to construct first the discontinuity (spectral)
function $\rho_1(\sigma) \equiv {\rm Im} \A_1(Q^2=-\sigma - i \epsilon)$ (for
$\sigma > 0$). This approach leads, in general, to nonperturbative
analytic QCD, i.e., $\A_n$ turns out to be different from $\A_1^n$. 
We can follow here analogous ideas used in the construction of
the spectral functions of spacelike observables (correlators)
in the literature, e.g., 
\cite{NestanAdl,DeRafael1,DeRafael2,Magradze:2010gn}.
In these references, analytization is applied
directly to a considered (spacelike) observable 
${\cal D}(Q^2)$ itself. Some of the new nonperturbative
parameters introduced there were thus specific to the chosen observable.
On the other hand, we take here the standpoint that it is the (universal)
QCD coupling that needs analytization; and that the additional 
nonperturbative contributions for a considered observable,
not contained in the analytized leading-twist contribution, 
are accounted for by a procedure containing other universal parameters.
Such parameters can be vacuum expectation values of higher
dimensional operators, and the aforementioned additional
nonperturbative contributions are represented by 
higher-twist terms of the Operator Product Expansion (OPE)
\cite{Shifman:1978bx}.

In our approach, at a large enough threshold value
$\sigma_0 = M_0^2$  ($\sim 1 \ {\rm GeV}^2$) we have the onset
of the perturbative behavior for the discontinuity function
of the coupling
\be
\rho_1(\sigma) = \rho_1^{\rm (pt)}(\sigma) \ , \ {\rm for} \
\sigma \geq M_0^2 \ .
\label{ponset}
\ee
On the other hand, in the regime $0 < \sigma < M_0^2$, 
the behavior is nonperturbative and unknown in detail,
and could be parametrized as a sum (with different weights) of
delta functions
\be
\rho_1(\sigma) = \sum_{j=1}^n F_n^2 \delta(\sigma - M_n^2) \ ,
\ {\rm for} \ 0 < \sigma < M_0^2 \ .
\label{deltas1}
\ee
As has been argued in \cite{Peris,CM}, introduction
of a sufficient number of positive delta functions in the
discontinuity function $\rho(\sigma) = {\rm Im} f(Q^2=-\sigma - i \epsilon)$
can approximate sufficiently well any positive Stieltjes function
$f(Q^2)$. The analytic coupling $\A_1(Q^2)$ 
is a positive Stieltjes function \cite{CM}. On the other hand,
application of such type of approximations has been applied to
the spectral functions of certain current correlators \cite{DeRafael2},
under the name of the ``minimal hadronic ansatz.'' 

In \cite{CCEM}, we constructed in this way a simple 
one-delta analytic QCD model, by introducing one delta function in the
low-$\sigma$ regime. The model contains three free parameters,\footnote{
In addition to the scale $\Lambda_{\rm QCD}$, 
which was fixed by the condition
of reproducing the world average value of $a_{\rm pt}(M_Z^2) = 0.119/\pi$,
in the ${\overline {\rm MS}}$ scheme.}
which were fixed by the condition (two requirements) 
$\A_1 - a_{\rm pt}  \sim (\Lambda^2/Q^2)^3$ at large $|Q^2| > \Lambda^2$
(where $\Lambda^2 = \Lambda^2_{\rm QCD} \sim 0.1 \ {\rm GeV}^2$), and the
(one) requirement of reproducing the correct value of the semihadronic
tau decay ratio $r_{\tau}$. 

It may be regarded as overly optimistic to approximate the unknown
low-$\sigma$ regime by a single delta funtion. In the present work, we
go beyond the one-delta approximation, and investigate how to parametrize
the low-$\sigma$ regime in a reasonable manner with two positive delta 
functions. Since such an extension introduces in the model several new
parameters, we will fix those parameters by specific reasonable
conditions, which will be similar in spirit to those of the
one-delta case. Specifically, the model can be made even
closer to pQCD, $\A_1 - a_{\rm pt}  \sim (\Lambda^2/Q^2)^5$ at large 
$|Q^2| > \Lambda^2$, while reproducing the correct value of $r_{\tau}$.

We do not introduce even more delta functions in
the low-$\sigma$ regime, because in such a case we would need values of
more low-energy QCD observables to fix at least some of the additional
parameters. Most of the inclusive low-energy
QCD observables, with the remarkable exception of the $(V+A)$
ratio $r_{\tau}$, either have large experimental uncertainties, or 
large theoretical uncertainties due to large higher-twist contributions,
or both. On the other hand, if we fix the parameters of the
model (with more than two deltas) by simply imposing
a further increase in the power index $n$ of the
difference  $\A_1 - a_{\rm pt}  \sim (\Lambda^2/Q^2)^n$ ($n>6$), and without
imposing the requirement of the reproduction of any additional 
low-energy observable value, some of the delta's become negative,
indicating numerical instabilities.

In Sec.~\ref{sec:descr} we describe the model and impose
the conditions which will fix the unknown parameters.
In Sec.~\ref{sec:rtau} we explain how to evaluate, in any
anQCD model, the leading-twist contribution of the
spacelike observables and of the related timelike observables,
among the latter being $r_{\tau}$.
In Sec.~\ref{sec:num} we present the numerical determination of the
model parameters and other numerical results. In Sec.~\ref{sec:summ}
we summarize the results of this work and outline the prospects of
further applications.

\section{Description of the two-delta model}
\label{sec:descr}

As outlined in the Introduction, we construct the two-delta
(2d) anQCD model by starting with an ansatz for the
discontinuity function 
$\rho_1(\sigma) \equiv {\rm Im} \  \A_1(Q^2=-\sigma - i \epsilon)$ (for $\sigma > 0$)
which agrees with the perturbative counterpart
$\rho_1^{\rm (pt)}(\sigma) \equiv {\rm Im} \  a_{\rm pt}(Q^2=-\sigma - i \epsilon)$
at sufficiently high scales 
$\sigma \geq M_0^2$ ($M_0^2 \sim 1 \ {\rm GeV}^2$);
and in the low-scale regime $0 < \sigma < M_0^2$ its otherwise
unknown behavior is parametrized as a linear combination of
(two) delta functions
\ba
\rho_1^{\rm (2d)}(\sigma; c_2) &=&
\pi \sum_{j=1}^2 f_j^2 \Lambda^2 \; 
\delta(\sigma - M_j^2) +  \Theta(\sigma-M_0^2) \times 
\rho_1^{\rm (pt)}(\sigma; c_2) 
\label{rho1o1}
\\
& = & \pi \sum_{j=1}^2 f_j^2 \; \delta(s - s_j) +  
\Theta(s-s_0) \times r_1^{\rm (pt)}(s) \ ,
\label{rho1o2}
\ea
where we denoted  $s=\sigma/\Lambda^2$, $s_j = M_j^2/\Lambda^2$ ($j=0,1,2$),
and $r_1^{\rm (pt)}(s; c_2) =  \rho_1^{\rm (pt)}(\sigma; c_2)
= {\rm Im} \ a_{\rm pt}(Q^2=-\sigma - i \epsilon; c_2)$. Here, 
$\Lambda^2$ ($\stackrel{<}{\sim} 10^{-1} \ {\rm GeV}^2$) is the Lambert scale
appearing in the following expression for $a_{\rm pt}$:
\ba
a_{\rm pt}(Q^2;c_2) = - \frac{1}{c_1} \frac{1}{\left[
1 - c_2/c_1^2 + W_{\mp 1}(z) \right]} \ ,
\label{aptexact}
\ea
where $Q^2=|Q^2| \exp(i \phi)$, the branches $W_{-1}$ and $W_{+1}$
of the Lambert function
refer to the case $0 \leq \phi < + \pi$ and $- \pi < \phi < 0$, 
respectively,\footnote{
In MATHEMATICA \cite{Math8}, the functions $W_n(z)$ are implemented
by the command ${\rm ProductLog}[n,z]$.}
and 
\be
z =  - \frac{1}{c_1 e} 
\left( \frac{|Q^2|}{\Lambda^2} \right)^{-\beta_0/c_1} 
\exp \left( - i {\beta_0}\phi/c_1 \right) \ ,
\label{zexpr}
\ee 
where the aforementioned Lambert scale $\Lambda^2$ appears.
The explicit expression (\ref{aptexact}) is the solution 
of the (perturbative) renormalization group equation (RGE)
of the form
\ba
\frac{\partial a_{\rm pt}(Q^2;c_2)}{\partial \ln Q^2} & = &
- \beta_0 a_{\rm pt}^2 \frac{\left[ 1 + (c_1 - (c_2/c_1)) a_{\rm pt}
\right]}{\left[ 1 - (c_2/c_1) a_{\rm pt} \right]} \ .
\label{RGE}
\ea
Here, $\beta_0 = (1/4) (11 - 2 n_f/3)$ and 
$c_1 = \beta_1/\beta_0 = (1/4) (102-38 n_f/3)/(11 - 2 n_f/3)$
are the universal constants, while $c_2 = \beta_2/\beta_0$ is the
free three-loop renormalization scheme parameter.
The expansion of the beta function $\beta(a_{\rm pt}) =
d a_{\rm pt}/d \ln Q^2$ in general gives
\be
\beta(a_{\rm pt}) = 
- \beta_0 a_{\rm pt}^2 
(1 + c_1 a_{\rm pt} + c_2 a_{\rm pt}^2 + c_3 a_{\rm pt}^3 + \ldots ) \ ,
\label{betapt}
\ee
where $c_j$ ($j\geq 2$) are general renormalization scheme parameters.
The choice of the beta function on the right hand side of (\ref{RGE})
gives $c_j = c_2^{j-1}/c_1^{j-2}$ ($j \geq 3$), which means that the three-loop
scheme parameter $c_2$ can be chosen freely in this form, 
while the higher-loop scheme
parameters are then fixed. The specific ``effective three-loop''
perturbative beta function of the Pad\'e form of Eq.~(\ref{RGE}) 
was chosen for 
convenience, because it gives an explicit solution (\ref{aptexact}), 
in terms of the branches of the Lambert function $W$ 
\cite{Gardi:1998qr,Magr1,Magr2}, and, at the same time, it allows us to vary
the renormalization scheme at the three-loop level ($c_2$).
In the following, it will turn out to be convenient to vary
the scheme parameter $c_2$ in the explicit solution, the
latter being used for 
$\rho_1^{\rm (pt)}(\sigma) \equiv {\rm Im} \  a_{\rm pt}(Q^2=-\sigma - i \epsilon)$
appearing in the discontinuity function (\ref{rho1o1})-(\ref{rho1o2}) 
of the anQCD model. 

\begin{figure}[htb] 
\begin{minipage}[b]{.49\linewidth}
\centering{\epsfig{file=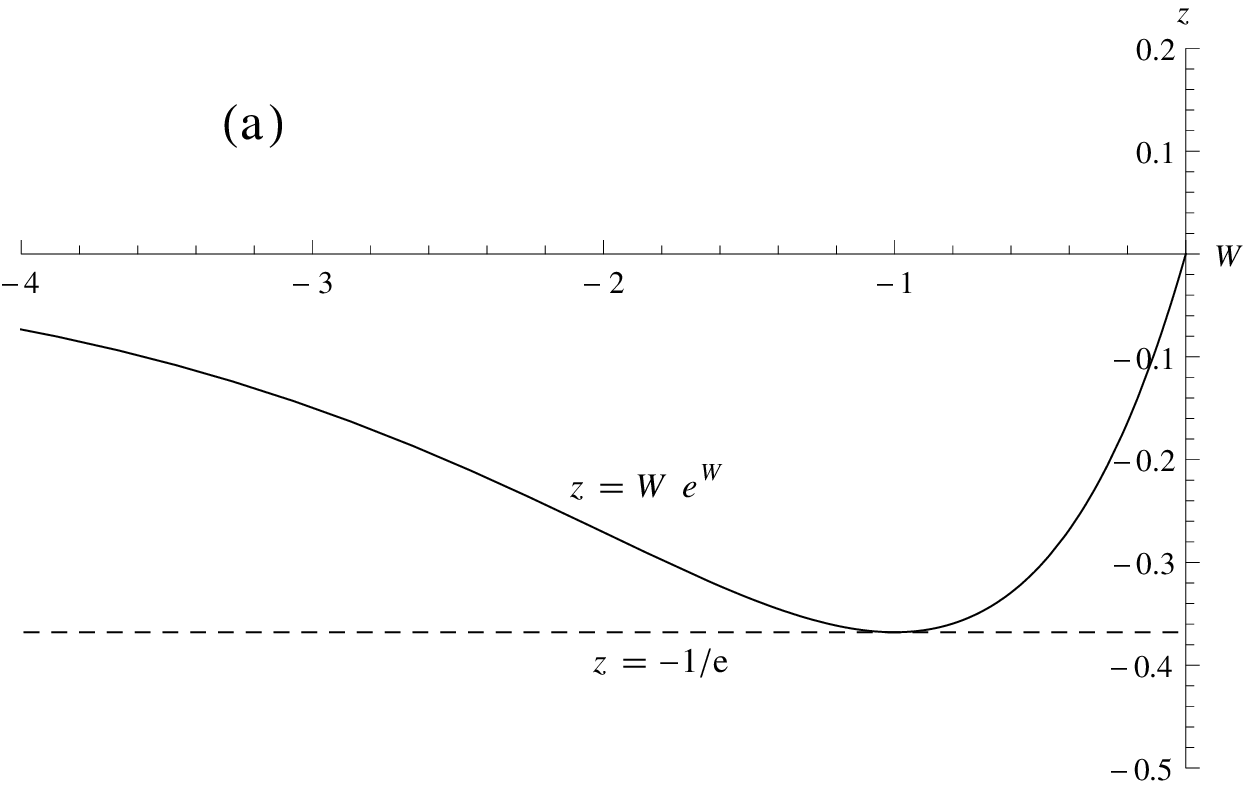,width=80mm,angle=0}}
\end{minipage}
\begin{minipage}[b]{.49\linewidth}
\centering{\epsfig{file=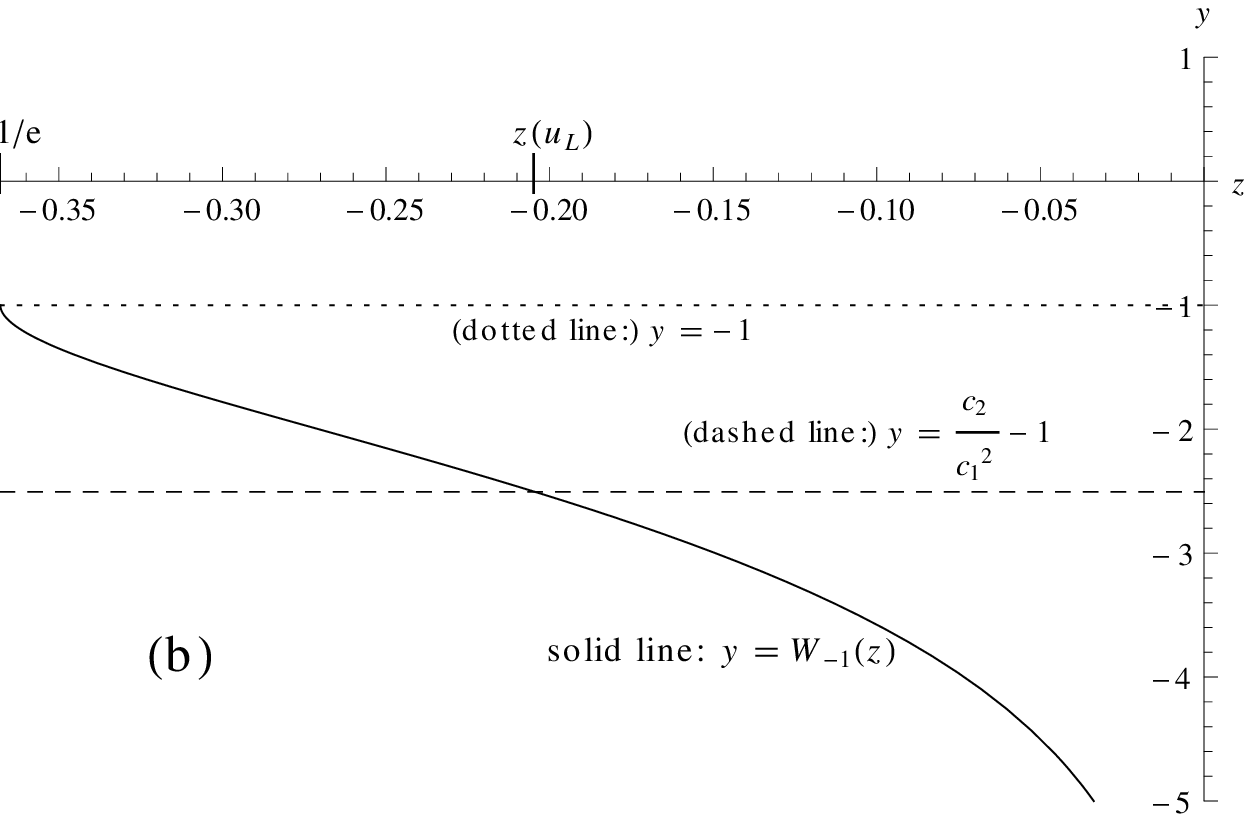,width=80mm,angle=0}}
\end{minipage}
\vspace{-0.4cm}
 \caption{\footnotesize (a) The defining relation $z = W e^W$ for
the Lambert function $W(z)$, for $-1/e < z < 0$; 
(b) The branch $W_{-1}(z)$ for the same $z$-interval; when $c_2<0$,
the denominator of Eq.~(\ref{aptexact}) becomes zero at a
$z(u_{\rm L})$ in this interval.}
\label{figACC1}
 \end{figure}

The Lambert function $W=W(z)$ is defined via the inverse relation
$z = W \exp(W)$, cf.~Fig.~\ref{figACC1}(a).
The two branches $W_{\mp 1}(z)$ of the Lambert function are related
via complex-conjugation $W_{+1}(z^{*}) = W_{-1}(z)^{*}$, 
and the point $z=-1/e$ is the branching point of these functions.
In the interval $-1/e < z < 0$, $W_{-1}(z)$ is a decreasing function of $z$,
cf.~Fig.~\ref{figACC1}(b). When $z \to -0$, the scale $Q^2$ tends
to $Q^2 \to + \infty$, cf.~Eq.~(\ref{zexpr}), and $W_{-1}(z) \to - \infty$,
this reflecting the asymptotic freedom of $a_{\rm pt}(Q^2)$ of
Eq.~(\ref{aptexact}). 
In our considered case of low-energy QCD 
(i.e., with number of quark flavors $n_f=3$), the solution 
(\ref{aptexact}) has unphysical (Landau) singularities along the positive 
$Q^2$ axis, for any $c_2$. An extension of such beta function to the effective
four- and five-loop case, such that the solution is explicit and
involving Lambert functions, was made in \cite{Cvetic:2011vy}.
For more details on the Lambert functions, we refer to 
Refs.~\cite{Corless:1996zz,Gardi:1998qr,Magr1,Magr2,Cvetic:2011vy}.

\begin{figure}[htb] 
\begin{minipage}[b]{.49\linewidth}
\centering{\epsfig{file=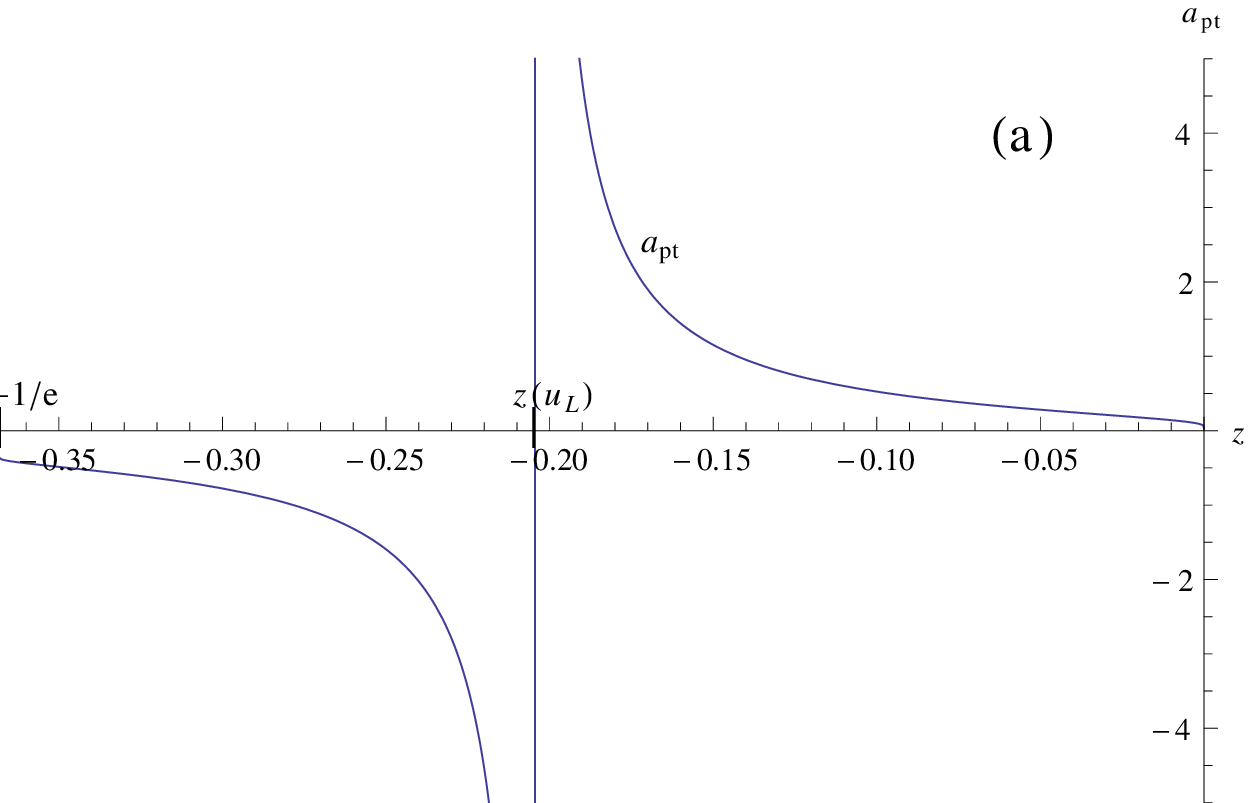,width=80mm,angle=0}}
\end{minipage}
\begin{minipage}[b]{.49\linewidth}
\centering{\epsfig{file=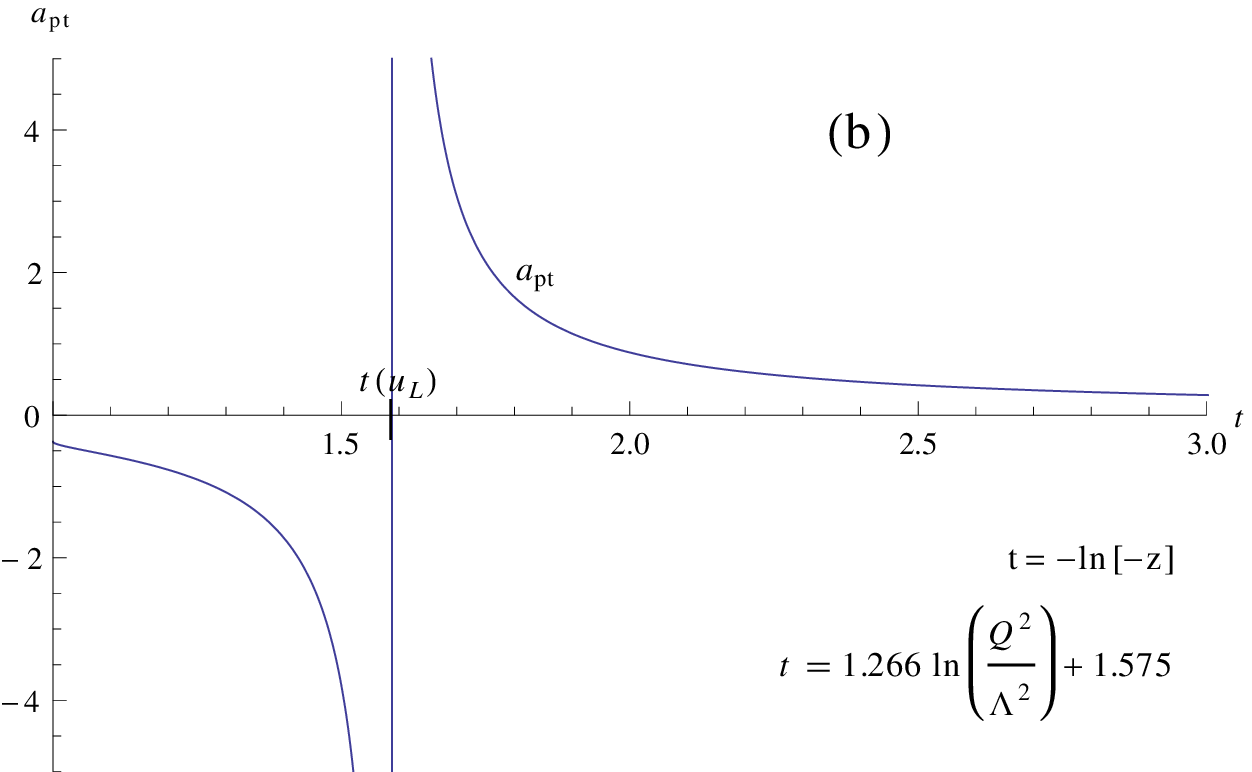,width=80mm,angle=0}}
\end{minipage}
\vspace{-0.4cm}
 \caption{\footnotesize (a) The perturbative coupling $a_{\rm pt}$
of Eq.~(\ref{aptexact}) as a function of $z$, for
$-1/e < z < 0$; (b) as a function of $t = - \ln(-z)$. The curves are for
the case of $n_f=3$ and $c_2=-4.76$ 
($\Rightarrow \ t = 1.266 \ln(Q^2/\Lambda^2) + 1.575$).}
\label{figACC2}
 \end{figure}

The aforementioned branching point of nonanalyticity 
$z(s_{\rm L})=-1/e$ corresponds, according to Eq.~(\ref{zexpr}), 
to the scale $Q^2(s_{\rm L}) = \Lambda^2 s_{\rm L}$ with
$s_{\rm L}=c_1^{-c_1/\beta_0}$ ($s_{\rm L} =0.6347$ when $n_f=3$), and the
interval $Q^2 \in (0, \Lambda^2 s_{\rm L})$ represents the interval
of the unphysical (Landau) singularities of $a_{\rm pt}(Q^2)$ of
Eq.~(\ref{aptexact}). If the scheme parameter $c_2$ is chosen to be negative
(this will be our case), then there is an additional pole-type
Landau singularity at a somewhat higher scale $Q^2(u_{\rm L}) =
\Lambda^2 u_{\rm L}$ ($\Leftrightarrow \ z=z(u_{\rm L}) =- u_{\rm L}^{-\beta_0/c_1}/(c_1 e)$)
at which the denominator of Eq.~(\ref{aptexact}) becomes zero,
cf.~Fig.~\ref{figACC1}(b), i.e., when
\be
-1 + c_2/c_1^2 = 
W_{-1}\left( \frac{-1}{c_1 e} |u_{\rm L}|^{-\beta_0/c_1} + i \epsilon \right) \ .
\label{uL}
\ee  
When $n_f=3$ and $c_2=-4.76$ (this will be our central choice of the scheme later),
we get $u_{\rm L} = 1.0095$ ($> s_{\rm L}$). For this case, the coupling
$a_{\rm pt}$ is presented in Fig.~\ref{figACC2}(a) as a function of
$z$ (for $-1/e < z < 0$, i.e., $s_{\rm L} \Lambda^2 < Q^2 < \infty$), 
and in Fig.~\ref{figACC2}(b) as a function of 
$t = - \ln(-z) = 1.266 \ln(Q^2/\Lambda^2) +  1.575$.

\begin{figure}[htb]
\centering{\epsfig{file=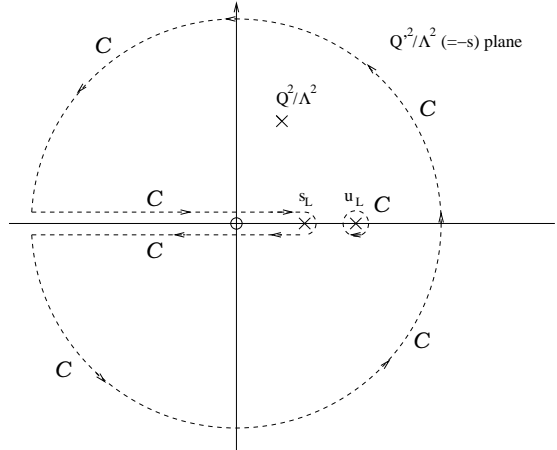,height=6.cm,angle=0}}
\vspace{-0.4cm}
\caption{The path of contour integration in $(Q^{' 2}/\Lambda^2)$-plane
leading to the expression (\ref{anptdispgen1}).}
\label{figACC3}
\end{figure}

It can be checked that, as a result of application of the Cauchy
theorem to the function $a_{\rm pt}(Q^{' 2})/(Q^{' 2} - Q^2)$ in the 
complex-$Q^{' 2}$ plane, the following dispersion relation for
$a_{\rm pt}$ holds:
\begin{equation}
a_{\rm pt}(Q^2;c_2) = 
\frac{1}{\pi} 
\int_{\sigma= - Q^2_{\rm min}-\eta^{'}}^{\infty}
d \sigma \; 
\frac{{\rm Im}  \; a_{\rm pt}(-\sigma-i \epsilon; c_2)}{(\sigma + Q^2)} 
=  \frac{1}{\pi} 
\int_{s= - u_{\rm min}-\eta}^{\infty} ds \; 
\frac{ r_1^{\rm (pt)}(s; c_2)}{(s + Q^2/\Lambda^2)} 
\qquad (\eta, \eta^{'} \to 0) \ ,
\label{anptdisp}
\end{equation}
where the integration covers the entire cut, i.e., starting
at a sufficiently low negative value $\sigma_{\rm min} = -Q^2_{\rm min}$ 
($Q^2_{\rm min} \stackrel{<}{\sim} 1 \ {\rm GeV}^2$). 
The perturbative discontinuity function is denoted as
$r_1^{\rm (pt)}(s;c_2) = {\rm Im} \ a_{\rm pt}(Q^2=-s \Lambda^2- i \epsilon; c_2)$.
Since the cut of the coupling $a_{\rm pt}(Q^{' 2},c_2)$, Eq.~(\ref{aptexact}) with
$c_2<0$, includes also the pole $Q^{' 2}_{\rm L} = u_{\rm L} \Lambda^2$ of the coupling,
the contour of integration in the complex $(Q^{' 2}/\Lambda^2)$-plane
is of the type as presented in Fig.~\ref{figACC3} (with the outer
radius going to infinity). 
Therefore,
the dispersive relation (\ref{anptdisp}) obtains a
slightly generalized form
\ba
a_{\rm pt}(Q^2) &=& 
\frac{1}{\pi} 
\int_{s= - s_{\rm L} - \eta}^{\infty} ds \;
\frac{r_1^{\rm (pt)}(s;c_2)}{(s + Q^2/\Lambda^2)} +
 \frac{\Delta u}{2 \pi} \int_{\Phi=-\pi}^{\pi}
d \Phi \ e^{ i \Phi} \; 
\frac{
a_{\rm pt} \left( (u_{\rm L} + \Delta u e^{i \Phi} ) \Lambda^2; c_2 \right)
}{\left[ (Q^2/\Lambda^2) - u_{\rm L} - \Delta u e^{i \Phi} \right]}
\label{anptdispgen1}
\\
& = &\frac{1}{\pi} 
\int_{s= - s_{\rm L} - \eta}^{\infty} ds \;
\frac{r_1^{\rm (pt)}(s;c_2)}{(s + Q^2/\Lambda^2)} + 
\frac{{\rm Res}_{(z=u_{\rm L})} a_{\rm pt}(z \Lambda^2; c_2)}{ (- u_{\rm L} + Q^2/\Lambda^2) } \ .
\label{anptdispgen2}
\ea
In Eq.~(\ref{anptdispgen1}), $\Delta u \Lambda^2$ is a sufficiently small 
(but otherwise arbitrary) radius
of integration around the point $u_{\rm L} \Lambda^2$ in the complex
$Q^{'2}$-plane, cf.~Fig.~\ref{figACC3}. 
In Eq.~(\ref{anptdispgen2}), this integration is
expressed by the residue of the function $a_{\rm pt}(z \Lambda^2)$ 
at this point.

The perturbative discontinuity function 
$r_1^{\rm (pt)}(s;c_2) = {\rm Im} \ a_{\rm pt}(Q^2=-s \Lambda^2- i \epsilon; c_2)$,
which is nonzero for $-s_{\rm L} < s < + \infty$, has the specific form
\ba
r_1^{\rm (pt)}(s;c_2) & = & 
{\rm Im} \left[ \frac{(-1)}{c_1} 
\frac{1}{\left[ 1 - (c_2/c_1^2) + W_{+1} \left( \frac{-1}{c_1 e} |s|^{-\beta_0/c_1} \exp( i \beta_0 \pi/c_1) \right) \right]} \right] 
\quad (s > 0) \ ,
\label{r1spos}
\\
& = &
{\rm Im} \left[ \frac{(-1)}{c_1} 
\frac{1}{\left[ 1 - (c_2/c_1^2) + W_{+1} \left( \frac{-1}{c_1 e} |s|^{-\beta_0/c_1} - i \epsilon \right) \right]} \right] 
\quad (s < 0) \ .
\label{r1sneg}
\ea

The analytic (spacelike) coupling $\A_1(Q^2;c_2)$ of
the two-delta anQCD model is constructed  on
the basis of the discontinuity function (\ref{rho1o2})
[cf.~Eq.~(\ref{r1spos})] using the dispersion relation (\ref{dispA1}).
This gives
\be
\A_1(Q^2;c_2) = \sum_{j=1}^2 \frac{f_j^2}{(u +s_j)} +
\frac{1}{\pi} \int_{s_0}^{\infty} ds \; 
\frac{r_1^{\rm (pt)}(s;c_2)}{(s+u)} \ ,
\label{A1Q2}
\ee
where $u = Q^2/\Lambda^2$. 

In the presented two-delta anQCD model, 
we will consider the first three quark flavors 
to be massless, and will consider that the momenta of the
$n_f=3$ regime in the anQCD model reach up to the threshold
$|Q^2| = (2 m_c)^2$ ($\approx 6.45 \ {\rm GeV}^2$).
Further, the anQCD model will be constructed in such a way as to
practically merge with pQCD at such sufficiently high momenta
[in the considered renormalization
scheme as fixed by Eq.~(\ref{RGE})].
Therefore, we will consider that the value
of the Lambert scale $\Lambda^2$ 
used in our analytic coupling $\A_1(Q^2;c_2)$ 
coincides with the perturbative 
Lambert scale $\Lambda_{\rm pt}^2$, the latter being determined
by the condition 
$a_{\rm pt}^{({\overline {\rm MS}})}(M_Z^2) = 0.1184/\pi$, i.e., by the
central value of the world average \cite{PDG2010}.
Therefore, $\Lambda^2$ is determined by
RGE-evolving this $a_{\rm pt}$ 
down to $a_{\rm pt}^{({\overline {\rm MS}})}((2 m_c)^2; n_f\!=\!3)$,
using the four-loop polynomial form of 
$\beta^{({\overline {\rm MS}})}(a_{\rm pt})$, and the three-loop matching
conditions \cite{CKS} at quark thresholds  
$\mu^2=(2 m_q)^2$ ($q=b,c$); and then changing from the
${\overline {\rm MS}}$ scheme to the scheme 
'$c_2$' [$\equiv (c_2, c_3=c_2^2/c_1, \ldots)$]
defined by the beta function of Eq.~(\ref{RGE}) 
(as explained, e.g., in Refs.~\cite{CKV,CCEM}).
 
The conditions we impose to fix the parameters are the following:
\begin{enumerate}
\item
The analytic coupling $\A_1(Q^2;c_2)$ acquires the aforementioned
pQCD value of the scale $\Lambda^2$ of 
$a_{\rm pt}(Q^2; c_2; n_f=3)$ at $Q^2 = (2 m_c)^2$
\be
\Lambda^2 = \Lambda^2_{\rm pt}(n_f=3) \ .
\label{pQCDc}
\ee
\item
While in general we expect $\A_1(Q^2;c_2)$ to differ from
$a_{\rm pt}(Q^2;c_2)$ at $Q^2 > \Lambda^2$ by $\sim (\Lambda^2/Q^2)^1$,
we impose the condition 
\be
\A_1(Q^2;c_2) - a_{\rm pt}(Q^2;c_2) \sim 
(\Lambda^2/Q^2)^{n_{\rm max}} \quad {\rm with} \; n_{\rm max}=5 \ .
\label{ITEPc}
\ee
\end{enumerate}
The condition (\ref{ITEPc}) represents in practice four conditions.
In the following few lines we will describe how to
formulate these four conditions. 

When we subtract from the perturbative
coupling (\ref{anptdispgen2}) the analytic coupling (\ref{A1Q2}),
we obtain
\be
a_{\rm pt}(Q^2;c_2) - \A_1(Q^2;c_2) =
\frac{{\rm Res}_{(z=u_{\rm L})} a_{\rm pt}(z \Lambda^2)}{ (u- u_{\rm L}) }
- \sum_{j=1}^2 \frac{f_j^2}{(u +s_j)} + \frac{1}{\pi}
\int_{s_{\rm L} - \eta}^{s_0} ds \; \frac{r_1^{\rm (pt)}(s;c_2)}{(s+u)} \ .
\label{diff}
\ee
Expanding the left-hand side in powers of $(1/u) = (\Lambda^2/Q^2)$,
the imposition of the condition (\ref{ITEPc}) gives us
the conditions that the terms of $\sim (\Lambda^2/Q^2)^{1+k}$
($k=0,1,2,3,$) in this expansion give us zero, i.e.,
we have the following four conditions
\ba
 \frac{1}{\pi}
\int_{s_{\rm L} - \eta}^{s_0} ds \; s^k \; r_1^{\rm (pt)}(s;c_2) +
(-u_{\rm L})^k \; 
{\rm Res}_{(z=u_{\rm L})} a_{\rm pt}(z \Lambda^2) & = &
s_1^k f_1^2 + s_2^k f_2^2 \quad (k=0,1,2,3) \ .
\label{ITEP4c}
\ea

Altogether, Eqs.~(\ref{pQCDc}) and (\ref{ITEP4c}) represent
five conditions. Once the
scheme $c_2$ parameter is chosen, we have altogether six
parameters in the model: $f_1^2$, $f_2^2$, $s_1$, $s_2$, $s_0$, 
and the scale $\Lambda$. Therefore, yet another condition
will have to be imposed, possibly involving the correct
reproduction of a low-energy observable. The best candidate for this
appears to be the canonical $(V+A)$ nonstrange and massless
semihadronic  $\tau$-lepton decay ratio $r_{\tau}$, 
\cite{ALEPH1,ALEPH2,OPAL}.\footnote{
The ``canonical'' means that the normalization is used such
that $(r_{\tau})_{\rm pt} = a_{\rm pt} + {\cal O}(a_{\rm pt}^2)$.}
When we remove the (measured) strangeness-changing
contribution, the color and CKM factors and
the electroweak effects, and the chirality-violating higher
twist (quark mass) contributions, the following value is 
obtained (cf.~\cite{DDHMZ,Ioffe,MY}, and
Appendix B of Ref.~\cite{CKV} for details)
\be
r_{\tau}(\triangle S=0, m_q=0)_{\rm exp.} =
0.203 \pm 0.004 \ .
\label{rtauexp}
\ee 
Numerical analyses of the measured data indicate that
the chirality-conserving higher-twist effects,
such as gluon-condensate contributions, are
negligible in the case of the considered $V+A$ decay channel.
Although such analyses have been performed
within pQCD+OPE approach, we will assume that they remain
valid when an analysis is performed with the
presented anQCD two-delta model + OPE. This assumption
appears to be reasonable because the considered anQCD coupling $\A_1(Q^2)$
is very close to the pQCD coupling $a_{\rm pt}(Q^2)$ (in the considered scheme)
at momenta $|Q^2| \stackrel{>}{\sim} 1 \ {\rm GeV}^2$. Therefore, in the
calculation of the discussed $r_{\tau}$, Eq.~(\ref{rtauexp}),
in the presented anQCD model, the (chirality-conserving)
higher-twist contributions will be ignored.

\section{Calculation of Adler function and $r_{\tau}$ in analytic QCD}
\label{sec:rtau}

The calculation of $r_{\tau}$ is then performed in the
same way as presented in \cite{CKV,CCEM}, i.e., by
performing explicitly the integration corresponding to the
leading-$\beta_0$ (LB) resummation for $r_{\tau}$, and adding the
three known beyond-LB (bLB) terms (i.e., including the exact
contributions of $\sim \A_4$). 

In this Section we will present only the main points
of calculation of spacelike (such as Adler function)
and timelike quantities (such as $r_{\tau}$) in
anQCD models. For details,
we refer to \cite{CV1,CV2,CCEM}, and
especially Appendices C and D of \cite{CKV}.

The basic idea of the approach in the evaluation
of spacelike observables ${\cal D}(Q^2)$ in general
anQCD model is to reorganize first the perturbation
series ${\cal D}(Q^2)_{\rm pt}$
\be
{\cal D}(Q^2)_{\rm pt} = a_{\rm pt} +
d_1 a_{\rm pt}^2 + d_2 a_{\rm pt}^3 +\ldots \ ,
\label{pt}
\ee
into the modified perturbation series (mpt)
\be
{\cal D}(Q^2)_{\rm mpt} = a_{\rm pt} 
+ {\widetilde d}_1 {\widetilde a}_{{\rm pt},2}  
+ {\widetilde d}_2 {\widetilde a}_{{\rm pt},3} + 
\ldots \ ,
\label{mpt}
\ee
where ${\widetilde d}_k {\widetilde a}_{{\rm pt},k+1}$ are
the logarithmic derivatives of $a_{\rm pt}$
\be
{\ta}_{{\rm pt},k+1}(Q^2)
\equiv \frac{(-1)^{k}}{\beta_0^{k} k!}
\frac{ \partial^k a_{\rm pt}(Q^2)}{\partial (\ln Q^2)^k} \ , 
\qquad (k=0,1,2,\ldots) \ ,
\label{tan}
\ee
They are related with the powers of $a_{\rm pt} \equiv \alpha_s/\pi$ 
in the following way [using renormalization group equations (RGE) in pQCD]:
\ba
{\widetilde a}_{{\rm pt},2} &=&
a_{\rm pt}^2 + c_1 a_{\rm pt}^3 + c_2 a_{\rm pt}^4 + \ldots \ ,
\label{ta2}
\\
{\widetilde a}_{{\rm pt},3} &=&
a_{\rm pt}^3 + \frac{5}{2} c_1 a_{\rm pt}^4 + \ldots \ ,
\quad 
{\widetilde a}_{{\rm pt},4} =
a_{\rm pt}^4 + \ldots \ , \qquad {\rm etc.}
\label{ta34}
\ea 
This, in turn, means that the powers of $a_{\rm pt}$ are linear combinations
of logarithmic derivatives
\ba
a_{\rm pt}^2 & = & {\widetilde a}_{{\rm pt},2}
- c_1 {\widetilde a}_{{\rm pt},3} + \left( \frac{5}{2} c_1^2 - c_2 \right)
{\tilde a}_{{\rm pt},4} + \ldots \ ,
\label{a2}
\\
a_{\rm pt}^3 & = & {\widetilde a}_{{\rm pt},3}
- \frac{5}{2} c_1 {\widetilde a}_{{\rm pt},4} + \ldots \ ,
\qquad
 a_{\rm pt}^4  =  {\widetilde a}_{{\rm pt},4} + \ldots \ , 
\qquad {\rm etc.} \ ,
\label{a34}
\ea 
which allows us to relate the mpt coefficients with
the usual perturbation series (pt) coefficients
\ba
{\widetilde d}_1 & = & d_1 \ , \qquad
{\widetilde d}_2 =  d_2 - c_1 d_1 \ ,
\label{td12}
\\
{\widetilde d}_3 & = & d_3 - \frac{5}{2} c_1 d_2 + 
\left( \frac{5}{2} c_1^2 - c_2 \right) d_1 \ ,
\qquad {\rm etc.}
\label{td3}
\ea
In \cite{CV1,CV2} it was shown that it is imperative to construct 
first the analogs of the logarithmic derivatives of $a_{\rm pt}$ in 
the following way:\footnote{
If the analytization is performed in any other way, the renormalization scale and
scheme dependence of the resulting truncated analytic series of any observable
${\cal D}(Q^2)$ will in general increase (instead of decrease) when the number 
of terms in the series increases, cf.~\cite{CV1,CV2}.}
\be
\left( \frac{\partial^k a_{\rm pt}(Q^2)}
{\partial (\ln Q^2)^k} \right)_{\rm an}
=
\frac{\partial^k}{\partial (\ln Q^2)^k} 
\left( a_{\rm pt}(Q^2) \right)_{\rm an} 
= 
\frac{\partial^k \A_1(Q^2)}{\partial (\ln Q^2)^k} \qquad (k=0,1,2,\ldots) \ .
\label{logder}
\ee
This means that the 'mpt' expansion (\ref{mpt}) becomes in anQCD
the corresponding ``modified analytic series'' ('man')
\be
{\cal D}(Q^2)_{\rm man} = \A_1 +
{\widetilde d}_1 \tA_2 + {\widetilde d}_2 \tA_3 + \ldots \ ,
\label{man}
\ee
and its truncated version is
\be
{\cal D}(Q^2)_{\rm man}^{[N]} = \A_1 +
{\widetilde d}_1 \tA_2 + \ldots + 
{\widetilde d}_{N-1} \tA_N \ .
\label{manN}
\ee
Here we denoted by $\tA_{k+1}$ the logarithmic derivatives
of $\A_1$
\be
\tA_{k+1}(\mu^2)
= \frac{(-1)^k}{\beta_0^k k!}
\frac{ \partial^k \A_1(\mu^2)}{\partial (\ln \mu^2)^k} \ ,
\qquad (k=1,2,\ldots) \ .
\label{tAn}
\ee
The expressions (\ref{man})-(\ref{tAn}) are the basis of the
evaluation of massless spacelike observables in any anQCD.\footnote{
If masses are involved in the evaluated physical quantity, 
perturbation series contains noninteger powers $a^{\nu}_{\rm pt}$
(and possibly derivatives thereof with respect to $\nu$, i.e.,
$a^{\nu}_{\rm pt} ln^k a_{\rm pt}$). 
The evaluation of such quantities in anQCD models
is then based on the procedure presented in \cite{GCAK}.}

The quantity $r_{\tau}$ is, on the other hand, a timelike
observable. However, it is expressed via a contour integration 
\cite{Braaten:1988hc} 
\be
r_{\tau} = \frac{1}{2 \pi} \int_{-\pi}^{+ \pi}
d \phi \ (1 + e^{i \phi})^3 (1 - e^{i \phi}) \
d_{\rm Adl} (Q^2=m_{\tau}^2 e^{i \phi}) \ ,
\label{rtaucont}
\ee
which
involves the (spacelike, massless, $n_f=3$) Adler function
$d_{\rm Adl}(Q^2) =   a_{\rm pt}(Q^2) + {\cal O}(a_{\rm pt}^2)$. 
The perturbation expansion of $d_{\rm Adl}$ is known
up to $\sim a_{\rm pt}^4$ \cite{d1,d2,d3}
\be
d_{\rm Adl} (Q^2)^{[4]}_{\rm pt} = a_{\rm pt} + 
\sum_{n=1}^{3} (d_{\rm Adl})_n a_{\rm pt}^{n+1} \ .
\label{dAdlpt}
\ee
On the other hand, the leading-$\beta_0$ parts 
$(d_{\rm Adl})_n^{\rm (LB)} = c_{n,n}^{(1)} \beta_0^n$  
[$= ( {\widetilde d}_{\rm Adl} )_n^{\rm (LB)}$] of
all the coefficients $(d_{\rm Adl})_n$ are known\footnote{We have 
$(d_{\rm Adl})_n =   c_{n,n}^{(1)} \beta_0^n + {\cal O}(\beta_0^{n-1})$ and
$( {\widetilde d}_{\rm Adl} )_n =  c_{n,n}^{(1)} \beta_0^n + {\cal O}(\beta_0^{n-1})$.
The expansions in powers of $\beta_0$ are obtained when $(d_{\rm Adl})_n$ and
$( {\widetilde d}_{\rm Adl} )_n$ are written in powers of $n_f$
($= - 6 \beta_0 +33/2$) and then reorganized in powers of $\beta_0$.}
\cite{Broadhurst,Beneke} 
and the resummation of these contributions can be performed
by using the approach of Neubert of integration with characteristic 
functions \cite{Neubert} - this can be performed in any anQCD 
without ambiguities (since no Landau singularities)
 \ba
(d_{\rm Adl})^{\rm (LB)}_{\rm man}(Q^2) &\equiv&
 \A_1(Q^2) + c_{1,1}^{(1)} \beta_0 {\tA}_{2}(Q^2) +
\ldots +   c_{n,n}^{(1)} \beta_0^n {\tA}_{n+1}(Q^2) + \ldots
\label{DLBman}
\\
& = &
\int_0^{\infty} \frac{dt}{t} F_{\rm Adl}(t) 
\A_1(t Q^2 e^{-5/3}) \ .
\label{DLBintan}
\ea
Here, $F_{\rm Adl}(t)$
is the characteristic function of the Adler function,
whose explicit expression was obtained in \cite{Neubert}.

On the other hand, the full coefficients $(d_{\rm Adl})_n$
and $({\widetilde d}_{\rm Adl})_n$ are known only up to $n=3$
\cite{d1,d2,d3}. Therefore, the full Adler function can be evaluated
in anQCD by adding to the leading-$\beta_0$ (LB) contribution the
three known beyond-LB (bLB) terms
\ba
(d_{\rm Adl})_{\rm man}^{\rm (LB+bLB)}(Q^2)^{[4]} & = &
\int_0^\infty \frac{dt}{t}\: F_{\rm Adl}^{\cal {E}}(t) \: 
\A_1(t Q^2 e^{-5/3}) +
\sum_{n=1}^{3} (T_{\rm Adl})_n \tA_{n+1} \ ,
\label{LBbLBmanN}
\ea
where 
\be
(T_{\rm Adl})_n = ({\widetilde d}_{\rm Adl})_n - c^{(1)}_{nn} \beta_0^n
\label{Tn}
\ee
are the complete bLB parts ($\sim \beta_0^{n-1}$)
of the coefficients $({\widetilde d}_{\rm Adl})_n$.

Using the expression (\ref{LBbLBmanN}) in the contour integration
(\ref{rtaucont}) gives for $r_{\tau}$ 
\be
r_{\tau}^{\rm (LB+bLB),[4]} = r_{\tau}^{\rm (LB)} + 
\sum_{n=1}^{3} \ (T_{\rm Adl})_n I(\tA_{n+1}) \ ,
\label{rtLBbLBman4}
\ee
where $I(\tA_{n+1})$ are the contour integrals
given by
\be
I(\tA_{n+1}) =
\frac{1}{2 \pi} \int_{-\pi}^{+ \pi}
d \phi \ (1 + e^{i \phi})^3 (1 - e^{i \phi}) \
\tA_{n+1}(m_{\tau}^2 e^{i \phi}) \quad (n=1,2,3) \ ,
\label{IanC}
\ee 
and the LB part in (\ref{rtLBbLBman4}) is
a well-defined (in anQCD models) integral of the form
\be
r_{\tau}^{\rm (LB)} = 
\frac{1}{\pi} \int_0^\infty \frac{dt}{t}\: {\widetilde F}_{r}(t) \: 
\rho_1(t m_{\tau}^2 e^{-5/3}) \ ,
\label{LBrt2}
\ee
where ${\widetilde F}_{r}(t)$ was calculated in \cite{CKV}
from the Minkowskian characteristic function $F_{r}^{\cal {M}}(t)$
of \cite{Neubert2}. For details on ${\widetilde F}_{r}(t)$
we refer to App.~D of \cite{CKV}.\footnote{
In (\ref{rtLBbLBman4})-(\ref{IanC}) we used in the Adler function
the renormalization scale $\mu^2=Q^2$ ($\equiv m_{\tau}^2 e^{i \phi}$); 
this scale, of course, can be varied,
cf.~\cite{CKV,CCEM} for details.}

Thus, the more explicit form of the $r_{\tau}$-reproduction condition,
mentioned in the previous section, is 
\be
\left( r_{\tau}^{\rm (LB+bLB[4])} = \right)
\; r_{\tau}^{\rm (LB)} + 
\sum_{n=1}^{3} \ (T_{\rm Adl})_n I(\tA_{n+1}) = 0.203 \ .
\label{rtauc}
\ee
The six conditions,
(\ref{pQCDc}), (\ref{ITEP4c}) and (\ref{rtauc}),
then determine the six parameters of the model:
$f_j^2, f_2^2, s_1, s_2, s_0$ and the scale $\Lambda$.
This procedure can be performed once we have chosen
a value of the scheme parameter $c_2$ of Eq.~(\ref{RGE}).

\section{Numerical results}
\label{sec:num}

In this Section we present the numerical results for the
parameters of the model, obtained from the
imposition of the aforementioned six conditions.
The additional (implicit) conditions that we choose are
that the weights $f_1^2$ and $f_2^2$ are positive. This is
based on the fact that the discontinuity function $\rho_1(\sigma)$
is positive in any reasonable scheme of pQCD. Furthermore,
the condition of positivity of $\rho_1(\sigma)$ can be 
expressed also via the condition that the
Minkowskian coupling
\be
\tlA_1(\sigma) = \frac{1}{\pi} \int_{\sigma}^{\infty} d \sigma^{'}
\frac{\rho_1(\sigma^{'})}{\sigma^{'}} 
\label{MinA1}
\ee
is a monotonously decreasing function of scale $\sigma$. 

If we choose for the scheme parameter $c_2=0$, we obtain
from the six conditions, as a result, that the pQCD-onset
mass $M_0 = \sqrt{s_0} \Lambda$ is relatively high,
$M_0 \approx 1.7$ GeV. Increasing $c_2$ to positive values
increases $M_0$ further; e.g., for $c_2=c_2^{\overline {\rm MS}}/2$
($\approx 2.24$) we get $M_0 \approx 2.0$ GeV, etc.
On the other hand, decreasing $c_2$ to negative values,
we obtain smaller $M_0$.
We believe that the effective pQCD-onset scale $M_0$ should be 
significantly smaller than the mass of the $\tau$ lepton.
This turned out to be so in the one-delta anQCD model
of \cite{CCEM}, with $c_2=0$, where $M_0 \approx 1$ GeV was obtained. 
In general, we do not want to parametrize (via delta functions)
relatively small deviations from pQCD, i.e., those at
$\sqrt{\sigma} > 1.5$ GeV. On the other hand, by lowering the value of
$c_2$, we encounter at $c_2 < -8.0$ ($M_0 < 0.9$ GeV) negative values of
$s_2$, implying that the analyticity is lost.

Therefore, we will adjust the scheme parameter $c_2$ in such a 
way as to get $M_0 = 1.25 \pm 0.25$ GeV. The results are given in
Table \ref{t1}. We believe that all three choices of $M_0$
will give almost the same predictions for various physical observables.
The reason for this lies in the renormalization scheme
independence of pQCD results; and our model, although nonperturbative
and analytic at low momenta $|Q| \stackrel{<}{\sim} 1$ GeV, is
practically indistinguishable from pQCD at all higher momenta.
\begin{table}
\caption{The parameters of the considered two-delta anQCD model,
for the three chosen values of the pQCD-onset scale $M_0$: $1.0; 1.25; 1.50$ GeV. 
In addition, the results of the one-delta (1d) model of
\cite{CCEM} in the scheme $c_2=0$ 
are given (the last line). See the text for details.}
\label{t1}  
\begin{ruledtabular}
\begin{tabular}{l|llllllll}
$M_0$ [GeV] & $c_2=\beta_2/\beta_0$ & $\Lambda$ [GeV] & $s_0$ & $s_1$ & $f_1^2$ & $s_2$ & $f_2^2$ & $\A_1(0)$
\\ 
\hline
1.00 & -7.15 & 0.193 & 26.86 & 19.473 & 0.3637 & 0.3594 & 0.7808 & 2.29
\\
1.25 & -4.76 & 0.260 & 23.06 & 16.837 & 0.2713 & 0.8077 & 0.5409 & 0.776
\\
1.50 & -2.10 & 0.363 & 17.09 & 12.523 & 0.1815 & 0.7796 & 0.3462 & 0.544
\\
\hline
0.886 (1d) & 0 & 0.472 & 3.525 & 0.4755 & 0.2086 & $\cdots$ & $\cdots$ & 0.544
\end{tabular}
\end{ruledtabular}
\end{table}
It is interesting that the value of the coupling
$\A_1$ at $Q^2=0$, obtained in this model for $M_0 = 1.25$ GeV
($\A_1(0) \approx   0.8$) is not far from the value
in Ref.~\cite{DSE1} ($\A_1(0) \approx 8.9/N_c/\pi \approx 0.9$),
which was obtained from an analysis using Dyson-Schwinger equations
for the ghost and gluon sector under an assumption of regularity 
of the ghost-gluon vertex.

In Table \ref{t1} we included, in the last line, also the
results of the one-delta (1d) model of \cite{CCEM},
obtained in an analogous way, 
in the scheme $c_2=c_3=\ldots = 0$.\footnote{
In Ref.~\cite{CCEM}, the parameters differ a little from those
in Table \ref{t1}, 
because the world average value taken there was from the year 2008,
$a_{\rm pt}(M_Z^2;{\overline {\rm MS}}) = 0.1190/\pi$, \cite{PDG2008};
and because we imposed there the condition $\A_1((3 m_c)^2) =
a_{\rm pt}((3 m_c)^2;n_f=3)$ instead of the (numerically similar) 
condition (\ref{pQCDc}).}
In that model, though, the smaller number of parameters
led to less stringent conditions (\ref{ITEPc}),
namaly with $n_{\rm max}=3$ 
(cf.~also a similar model in \cite{Alekseev:2005he}).

\begin{figure}[htb] 
\begin{minipage}[b]{.49\linewidth}
\centering{\epsfig{file=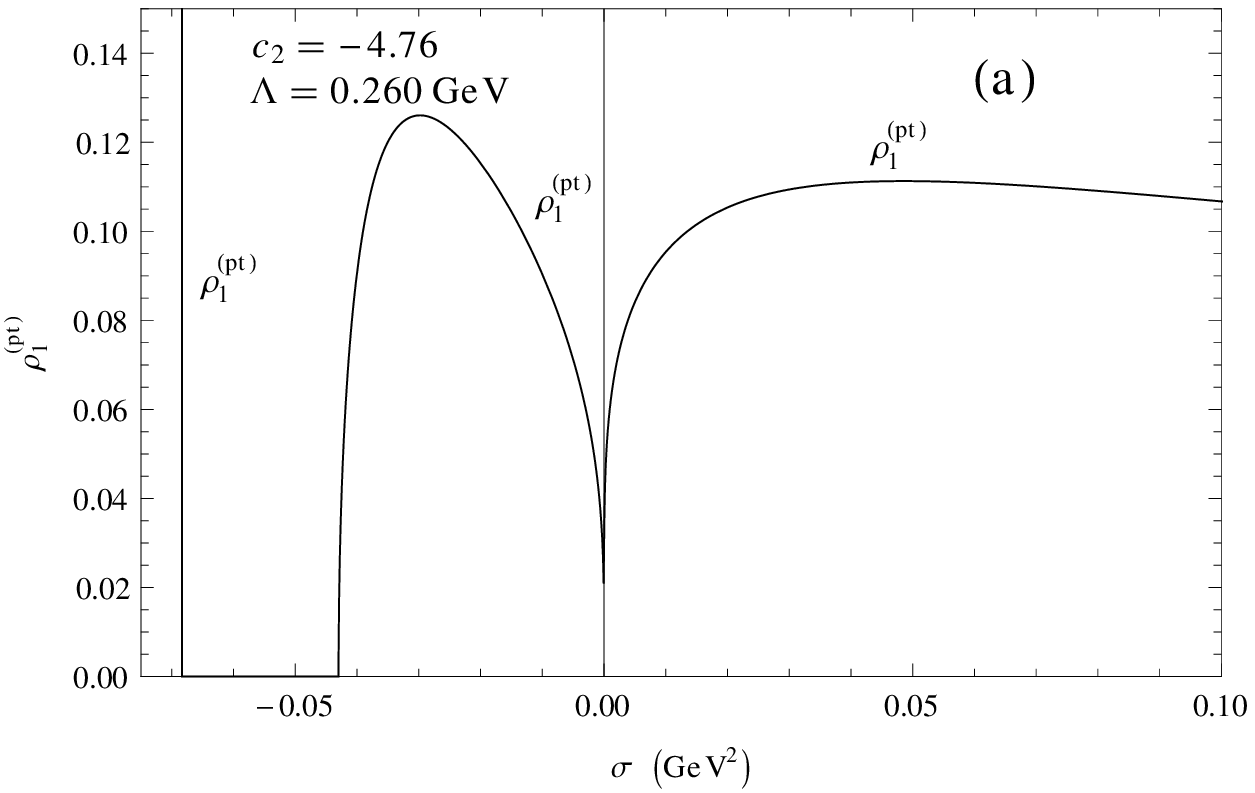,width=80mm,angle=0}}
\end{minipage}
\begin{minipage}[b]{.49\linewidth}
\centering{\epsfig{file=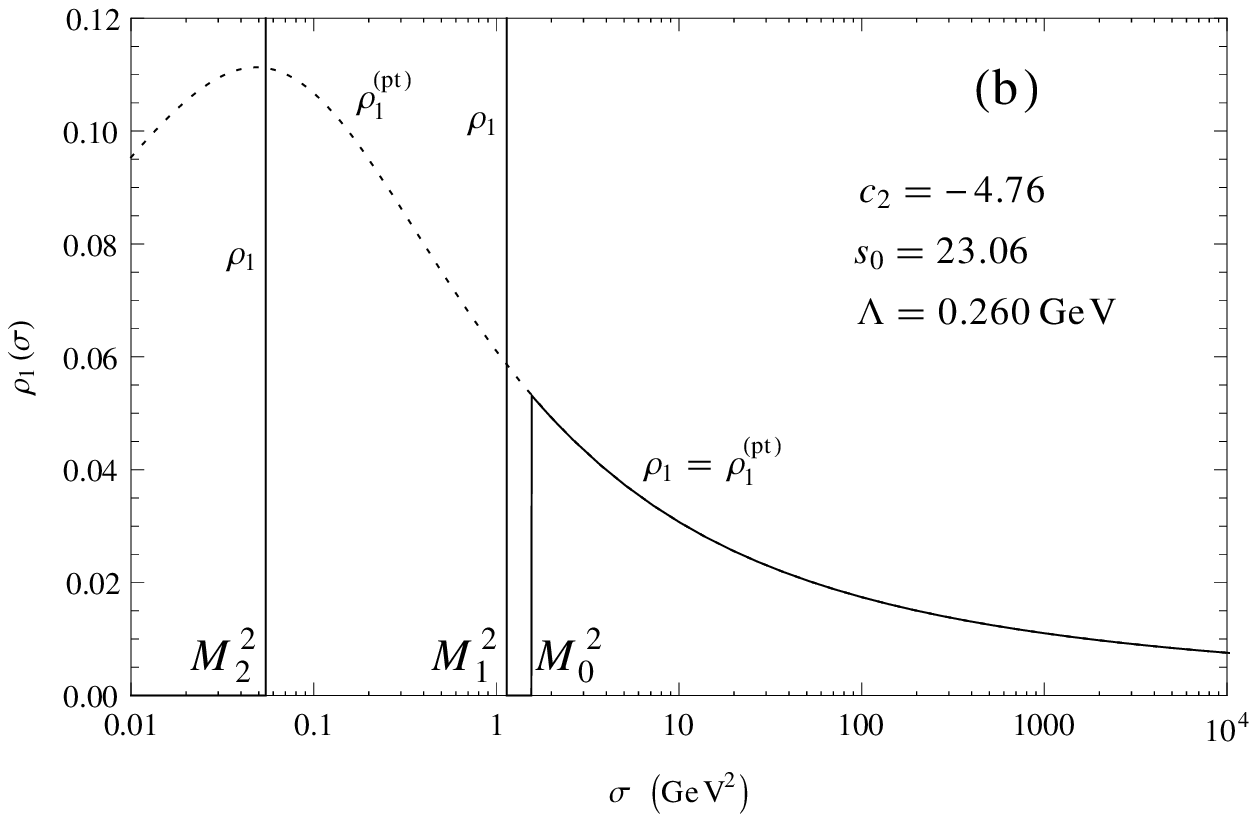,width=80mm,angle=0}}
\end{minipage}
\vspace{-0.4cm}
 \caption{\footnotesize The discontinuity function
$\rho_1(\sigma)$: (a) for the underlying pQCD coupling
$a_{\rm pt}(Q^2)$, where the Landau pole at $\sigma = - u_{\rm L} \Lambda^2$
($\approx - 0.068 \ {\rm GeV}^2$) and the branching point at
$\sigma = - s_{\rm L} \Lambda^2$ ($\approx - 0.043 \ {\rm GeV}^2$) are visible;
(b) for the considered two-delta
anQCD model, where the unknown region $0< \sigma < M_0^2$ is parametrized
by two delta functions, at $\sigma = M_1^2$ and $\sigma=M_2^2$. 
The parameters used correspond to the central case 
in Table \ref{t1} ($M_0=1.25$ GeV).}
\label{rho1Fig}
 \end{figure}

In Fig.~\ref{rho1Fig}(a), we present, for the resulting
''central'' choice of $M_0=1.25$ GeV ($c_2=-4.76$), 
the  corresponding pQCD discontinuity function
$\rho_1^{\rm (pt)}(\sigma) = {\rm Im} \ a_{\rm pt}(-\sigma - i \epsilon)$
of the underlying perturbative coupling $a_{\rm pt}$
(\ref{aptexact}), in the regime of low $|\sigma|$ and
including the unphysical (Landau) regime of
negative-$\sigma$ cut. 
We can note that in the latter regime
there is an additional, pole-like singularity at
$\sigma = - u_{\rm L} \Lambda^2$ ($\approx -0.261 \ {\rm GeV}^2$);
while the ``continuous'' part of $\rho_1^{\rm (pt)}(\sigma)$
ends a bit earlier, at 
$\sigma = - s_{\rm L} \Lambda^2 \approx - 0.207 \ {\rm GeV}^2$.\footnote{
Note: $s_{\rm L} = c_1^{-c_1/\beta_0}$, which is approximately
$0.635$ when $n_f=3$; at $Q^2 = s_{\rm L} \Lambda^2$ the Lambert
function $W_{-1}(z(Q^2))$ is equal to $-1$, and this is 
the branching point for $a_{\rm pt}(Q^2)$,
cf.~\cite{Gardi:1998qr,Cvetic:2011vy}.}
In Fig.~\ref{rho1Fig}(b),
$\rho_1(\sigma)$ of the considered two-delta anQCD
model is presented, cf Eq.~(\ref{rho1o1}).

\begin{figure}[htb] 
\begin{minipage}[b]{.49\linewidth}
\centering{\epsfig{file=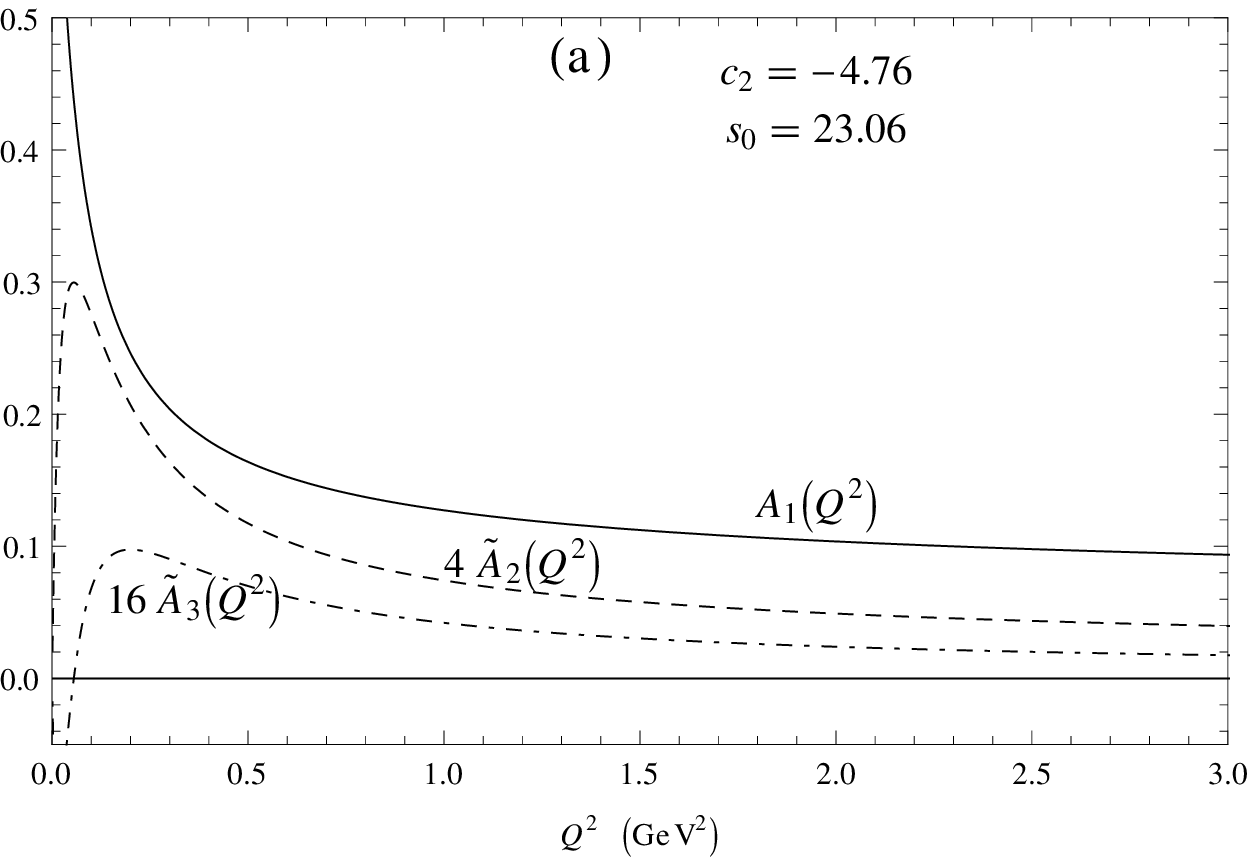,width=80mm,angle=0}}
\end{minipage}
\begin{minipage}[b]{.49\linewidth}
\centering{\epsfig{file=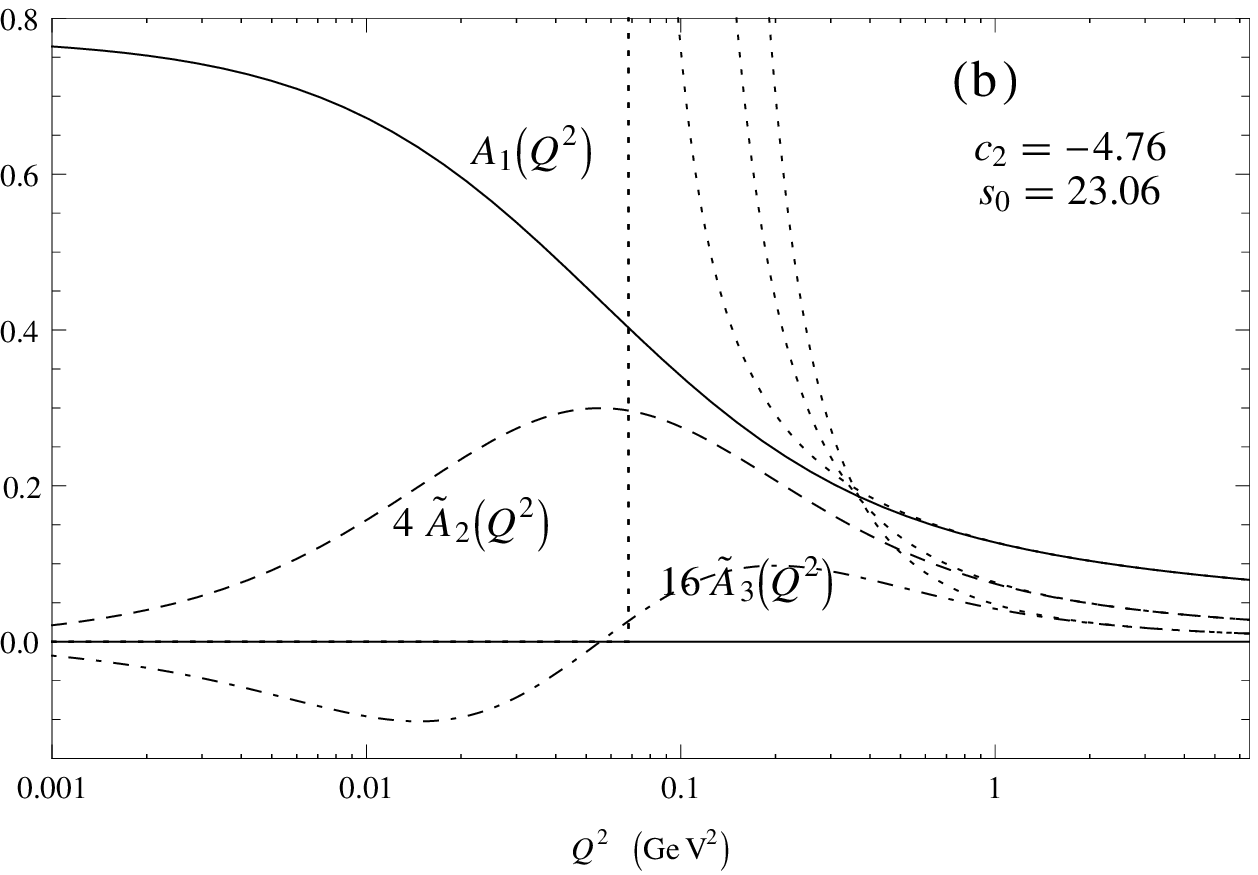,width=80mm,angle=0}}
\end{minipage}
\vspace{-0.4cm}
 \caption{\footnotesize The analytic couplings $\A_1$
(full line), $4 \times \tA_2$ (dashed curve), and 
$16 \times \tA_3$ (dash-dotted curve)
for positive $Q^2$, in the considered two-delta model, for the
central parameter choice of Table \ref{t1} ($M_0=1.25$ GeV): 
(a) linear scale is used for $Q^2$; 
(b) logarithmic scale is used for $Q^2$,
for better visibility at low values of $Q^2$. For comparison,
the corresponding pQCD couplings ($a_{\rm pt}$, 
$4 \times {\widetilde a}_{\rm pt,2}$, $16 \times {\widetilde a}_{\rm pt,3}$)
are included, as dotted curves (in the same renormalization scheme,
with $c_2=-4.76$); the vertical dotted line is the Landau pole
at $Q^2=u_{\rm L} \Lambda^2$ ($\approx 0.068 \ {\rm GeV}^2$).}
\label{couplsFig}
 \end{figure}

In Figs.~\ref{couplsFig}(a) and (b), we present, 
for the aforementioned central case $M_0=1.25$ GeV, the
resulting spacelike coupling $\A_1(Q^2)$ at positive $Q^2$.
The higher order couplings $\tA_k(Q^2)$ ($k=2,3$), 
cf.~Eq.~(\ref{tAn}), are also presented; due to a
strong hierarchy (even at low $Q^2$) they are rescaled,
for better visibility, by factors $4$ and $16$, respectively.
All the corresponding pQCD quantities ($a_{\rm pt}$, 
$4 {\widetilde a}_{{\rm pt},2}$, $16 {\widetilde a}_{{\rm pt},3}$) are
presented as dotted curves. For better visibility at low
$Q^2$, Fig.~\ref{couplsFig}(b) is presented with $Q^2$ on
logarithmic scale. It is clearly visible that the 
model practically agrees with the corresponding
pQCD model at $Q^2 > 1 \ {\rm GeV}^2$; and that for
$Q^2 < 1 \ {\rm GeV}^2$ the model differs from pQCD significantly,
due to the imposition of the analyticity.

\begin{figure}[htb] 
\begin{minipage}[b]{.49\linewidth}
\centering{\epsfig{file=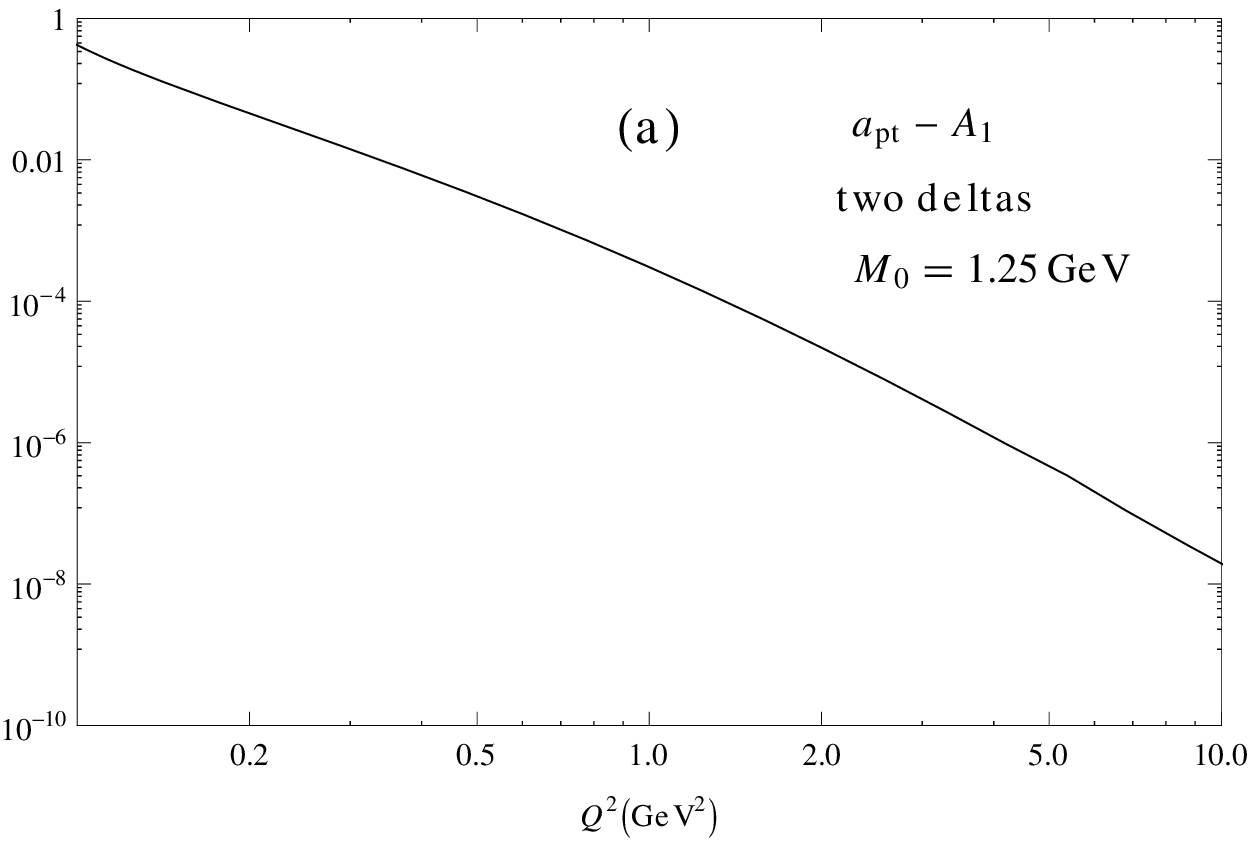,width=80mm,angle=0}}
\end{minipage}
\begin{minipage}[b]{.49\linewidth}
\centering{\epsfig{file=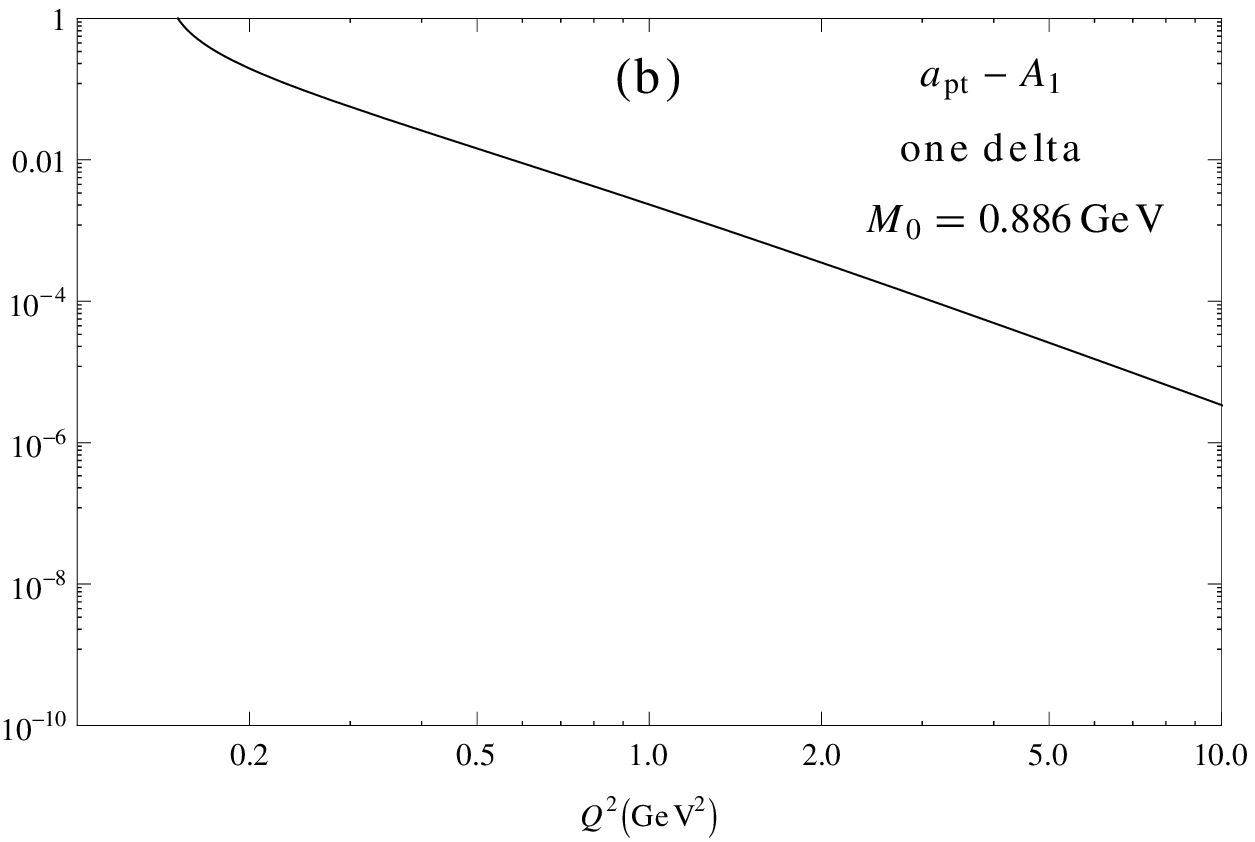,width=80mm,angle=0}}
\end{minipage}
\vspace{-0.4cm}
\caption{\footnotesize The difference between the perturbative and 
the analytic coupling, as a function of positive $Q^2$,
(a) the considered two-delta anQCD model (with $M_0=1.25$ GeV); 
(b) for the one-delta anQCD parameter of Ref.~\cite{CCEM}, but with 
the input parameters somewhat modified (see the text for details).}
\label{Figdiff}
 \end{figure}

In Fig.~\ref{Figdiff}(a), we present the difference between the
perturbative and the analytic coupling $(a_{\rm pt}(Q^2) - \A_1(Q^2))$ 
at positive $Q^2$, for the central choice 
of parameter ($M_0=1.25$ GeV) of Table \ref{t1}.
In the Figure we keep, formally, $n_f=3$ even for high $Q^2$. We see
that the difference vanishes fast when $Q^2$ increases.
For comparison, in Fig.~\ref{Figdiff}(b) we show the difference 
$(a_{\rm pt} - \A_1)$ in the case
of the one-delta anQCD model of \cite{CCEM}, with parameters as given in
the last line of Table \ref{t1}.

From Fig.~\ref{Figdiff}(a), we can deduce that the difference
$(a_{\rm pt} - \A_1)$ behaves as $\propto (\Lambda^2/Q^2)^{n_{\rm eff}}$,
where the numerical value of the effective power index
$n_{\rm eff}$ is somewhat less than 5; it is somewhere between
4 and 5. This can be understood in the following way: the
values of $s_0$ and $s_1$ in the model are relatively large
($\sim 10^1$), and therefore the coefficients ${\cal K}_{n}$
(for $n=5,6,...$) in the expansion
\be
a_{\rm pt}(Q^2;c_2) - \A_1(Q^2;c_2) = {\cal K}_5 (\Lambda^2/Q^2)^5 +
{\cal K}_6 (\Lambda^2/Q^2)^6 + \ldots 
\label{diff2}
\ee
are large ($\sim s_0^{n-1}, s_1^{n-1}$) and are increasing when $n$ increases.

Finally, in Table \ref{t2} we compare the convergence of the
\begin{table}
\caption{The four terms in the truncated analytic expansion
(\ref{rtLBbLBman4}) in two-delta anQCD model for the three
choices of the pQCD-onset scale $M_0$ and
for the one-delta anQCD model, as specified in Table \ref{t1}.}
\label{t2}  
\begin{ruledtabular}
\begin{tabular}{lll|llll|l}
$M_0$ [GeV] & $c_2$ & $\Lambda$ [GeV] & $r_{\tau}:$ LB & NLB & 
${\rm N}^2{\rm LB}$ &
${\rm N}^3{\rm LB}$ &
sum (sum) 
\\
\hline
1.00 & -7.15 & 0.193 & 0.2023 & 0.0012 & -0.0010 & 0.0005 & 0.2030 \\
1.25 & -4.76 & 0.260 & 0.2052 & 0.0013 & -0.0053 & 0.0019 & 0.2030 \\
1.50 & -2.10 & 0.363 & 0.2096 & 0.0013 & -0.0110 & 0.0030 & 0.2030 \\
\hline
0.886 & 0    & 0.472 & 0.2149 & 0.0014 & -0.0176 & 0.0043 & 0.2030
\end{tabular}
\end{ruledtabular}
\end{table}
series for $r_{\tau}$, Eq.~(\ref{rtLBbLBman4}), in the
three cases of the two-delta model and in the one-delta model
of Table \ref{t1}. We see that the convergence is good in general,
and appears to be better when the underlying pQCD scheme parameter 
$c_2$ is more negative.

There is one more aspect of the presented model that we should address.
We recall in more detail how we matched the presented anQCD model 
indirectly to the ${\overline {\rm MS}}$ scheme, via the condition 
(\ref{pQCDc}):
\begin{enumerate}
\item 
The Lambert scale of the model, at $n_f=3$, 
was chosen to coincide with the Lambert scale of the 
underlying pQCD model in the
same scheme determined by the beta function of Eq.~(\ref{RGE}) at $n_f=3$;
\item
The latter scale was fixed in such a way that the change of the 
scheme parameters $(c_2, c_2^2/c_1, c_2^3/c_1^2, \ldots)$ to the four-loop
${\overline {\rm MS}}$ scheme parameters 
$(c_2^{\overline {\rm MS}}, c_3^{\overline {\rm MS}})$, 
at $n_f=3$ and $\mu^2 = (2 m_c)^2$, gave the value of
$a_{\rm pt}^{({\overline {\rm MS}})}((2 m_c)^2;n_f=3)$ which corresponds to
$a_{\rm pt}^{({\overline {\rm MS}})}(M_Z^2) = 0.1184/\pi$, the
central value of the world average \cite{PDG2010}. The RGE-running
between $\mu^2=M_Z^2$ and $\mu^2=(2 m_c)^2$, in ${\overline {\rm MS}}$,
was performed in the usual way, using the four-loop polynomial form of 
$\beta^{({\overline {\rm MS}})}(a_{\rm pt})$, and the three-loop matching
conditions \cite{CKS} at quark thresholds  $\mu^2=(2 m_q)^2$ ($q=b,c$).
\end{enumerate}

As pointed out in Ref.~\cite{Binger:2003by},
the aforementioned (three-loop) matching in principle
introduces, indirectly, an element of nonanalyticity in the 
described framework, at\footnote{
We consider, throughout, the first three flavors to be massless.}
the scale $Q^2 = (2 m_c)^2$ and (to a much lesser degree) at $Q^2=(2 m_b)^2$.
This is so because the matching introduces nonanalyticity 
(even: discontinuity) in the running coupling 
$a_{\rm pt}^{({\overline {\rm MS}})}(Q^2)$ at those threshold scales.
It would be more convenient for the presented low-energy
anQCD model to be matched [at energies $Q^2 \sim (2 m_c)^2$] 
to a scheme which introduces the quark mass threshold effects in 
the running in a gradual (analytic) way. One such scheme is
the pinch technique (PT), Refs.~\cite{pinch}. The PT effective charge 
(i.e., running coupling) was presented for supersymmetric QCD
at one-loop in Ref.~\cite{Binger:2003by}. 
Specifically, in nonsupersymmetric QCD with three massless flavors, 
the relations of 
Ref.~\cite{Binger:2003by} (especially their Eq.~(A3)) imply
(for $|Q_0^2|, |Q^2| \alt (2 m_c)^2$)
\be
a_{\rm PT}(Q^2) = a_{\rm PT}(Q_0^2) + a_{\rm PT}^2(Q_0^2)
\left[ -\frac{9}{4} \ln(Q^2/Q_0^2) + \frac{1}{6} L_{1/2}(Q^2/m_c^2)
- \frac{1}{6} L_{1/2}(Q_0^2/m_c^2) \right] \ ,
\label{aPT}
\ee
where the quark threshold function $L_{1/2}$ is
\be
L_{1/2}(Q^2/m^2) = (3 - \beta^2) \left[ 
\beta \; {\rm ArcTanh}(1/\beta) - 1 \right] + 2 \ ,
\label{L12}
\ee
where $\beta = \sqrt{1 + 4 m^2/Q^2}$ and the (complex) momenta $Q^2$ are 
not on the cut $Q^2 < - 4 m^2$. In the case of decoupling ($|Q^2| \ll m^2$),
this function acquires the value of $(5/3)$.

We can estimate the practical errors introduced in the calculations
in our model due to the aforementioned matching to ${\overline {\rm MS}}$,
by estimating the errors that the actual nondecoupling of the charm
quark mass introduced in our calculation of $r_{\tau}$ (note that
in our calculation, we considered $c$ quark to be completely decoupled,
in the spirit of ${\overline {\rm MS}}$). The relation (\ref{aPT}) would
be transcribed in our model, approximately, in the following way:
\be
\A_1(Q^2)_{\rm thr} \approx \A_1(Q_0^2)_{\rm thr} + \tA_2({\bar Q}^2) 
\left[ -\frac{9}{4} \ln(Q^2/Q_0^2) + \frac{1}{6} L_{1/2}(Q^2/m_c^2)
- \frac{1}{6} L_{1/2}(Q_0^2/m_c^2) \right] \ ,
\label{A1thr}
\ee 
where $\tA_2({\bar Q}^2)$ is the logarithmic derivative of $A_1$ defined 
via Eq.~(\ref{tAn}), and the scale ${\bar Q}^2$ is taken to be
the geometric mean\footnote{
Formally, at one-loop level, it would be equivalent to take the
scale ${\bar Q}^2 = Q_0^2$ or ${\bar Q}^2 = Q^2$; numerically, though,
the geometric mean of these two scales (i.e., the arithmetic mean
of the logarithms of these scales) is better.} 
 of $Q_0^2$ and $Q^2$: ${\bar Q}^2 = \sqrt{Q_0^2 Q^2}$. Further, the
subscript ``thr'' indicates that the nondecoupling effect $m_c \not=
\infty$ is taken into account, in a first approximation. The corresponding
relation in the model without the analytic threshold effects is
\be
\A_1(Q^2) \approx \A_1(Q_0^2) + \tA_2({\bar Q}^2) 
\left[ -\frac{9}{4} \ln(Q^2/Q_0^2) \right] \ .
\label{A1nothr}
\ee 
An estimate of the error in the calculation of $r_{\tau}$, due to
the nondecoupling of $m_c$, in the calculation of $r_{\tau}$
can be obtained by using the relation (\ref{rtaucont}) in the
leading order (LO) when $d_{\rm Adl}(Q^2) = \A_1(Q^2)$ there, and subtracting
for $\A_1(Q^2)$ the two relations (\ref{A1thr}) and (\ref{A1nothr})
on the contour
\ba
\delta r_{\tau}^{\rm (LO)}(m_c \not= \infty) &=& 
\frac{1}{2 \pi} \int_{-\pi}^{+ \pi}
d \phi \ (1 + e^{i \phi})^3 (1 - e^{i \phi}) \
\left[ \A_1(Q^2)_{\rm thr} -  \A_1(Q^2) \right]
{\big |}_{Q^2=m_{\tau}^2 e^{i \phi}} 
\nonumber\\
& \approx &
\frac{1}{2 \pi} \int_{-\pi}^{+ \pi}
d \phi \ (1 + e^{i \phi})^3 (1 - e^{i \phi}) 
\nonumber\\
&& \times 
\left[ \left( \A_1(Q_0^2)_{\rm thr} -  \A_1(Q_0^2) \right)
+ \frac{1}{6} \tA_2(\sqrt{Q_0^2 Q^2}) 
\left(L_{1/2}(Q^2/m_c^2) - L_{1/2}(Q_0^2/m_c^2) \right) \right]
{\big |}_{Q^2=m_{\tau}^2 e^{i \phi}} .
\label{rtauthr1}
\ea
The scale $Q_0^2$ here should have a low value ($|Q_0^2| \ll m_{\tau}^2$)
so that the charm quark can be assumed to basically decouple
($m_c = \infty$) at such scale. Namely, in such a case, the first
(unknown) term in the integrand, $(\A_1(Q_0^2)_{\rm thr} -  \A_1(Q_0^2))$, can
be taken to be zero, and the remaining term would indicate correctly
the effects of nondecoupling of $m_c$. On the other hand, $|Q_0|^2$ cannot
be taken to be too small, as then the $\sim \tA_2$ term in the relations
 (\ref{A1thr}) and (\ref{A1nothr}) would dominate over the $\A_1(Q_0^2)$
term, due to a very large value of the logarithm $\ln (Q^2/Q_0^2)$. Therefore,
we will choose $Q_0^2 = \kappa Q^2$ ($=\kappa m_{\tau}^2 \exp(i \phi)$), with
$\kappa < 1$ varying in a specific interval were both aforementioned 
restrictions are reasonably fulfilled
\be
\delta r_{\tau}^{\rm (LO)}(m_c \not= \infty) \approx 
\frac{1}{12 \pi} \int_{-\pi}^{+ \pi}
d \phi \ (1 + e^{i \phi})^3 (1 - e^{i \phi}) 
\tA_2(\sqrt{Q_0^2 Q^2}) 
\left( L_{1/2}(Q^2/m_c^2) - L_{1/2}(Q_0^2/m_c^2) \right)
{\big |}_{Q^2=m_{\tau}^2 e^{i \phi}; Q_0^2= \kappa Q^2} .
\label{rtauthr2}
\ee
The results of these estimates, for the central value of the 
parameters of the presented two-delta anQCD model (i.e., the
case $M_0 = 1.25$ GeV of Table \ref{t1}) are presented in Table \ref{t3}.
\begin{table}
\caption{The estimates of nondecoupling of $m_c$ in the calculation
of $r_{\tau}$, Eq.~(\ref{rtauthr2}), for different value of the
parameter $\kappa$. In the third column we include a measure of the effect of 
nondecoupling of $m_c$ at $Q_0^2=\kappa m_{\tau}^2$ (assumed to be
negligible in Eq.~(\ref{rtauthr2})); and in the fourth column the ratio of
$\frac{9}{4} \ln(Q^2/Q_0^2) \tA_2({\bar Q}^2)$ with $\A_1(Q_0^2)$
(assumed to be small in (\ref{A1thr}) and (\ref{A1nothr})),
for $Q^2=m_{\tau}^2$ and $Q_0^2 =\kappa m_{\tau}^2$.}
\label{t3}  
\begin{ruledtabular}
\begin{tabular}{llll}
$\kappa$ & $\delta r_{\tau}^{\rm (LO)}(m_c \not= \infty)$ & $(L_{1/2}(\kappa m_{\tau}^2/m_c^2) - L_{1/2}(0))/L_{1/2}(0)$ & $ -\frac{9}{4} \ln(\kappa) \tA_2(\sqrt{\kappa} m_{\tau}^2)/\A_1(\kappa m_{\tau}^2)$
\\
\hline
%
0.15 & $6.1 \times 10^{-4}$ & 0.034 & 0.42 \\
0.20 & $4.6 \times 10^{-4}$ & 0.045 & 0.36 \\
0.30 & $3.1 \times 10^{-4}$ & 0.066 & 0.28 \\
0.40 & $2.3 \times 10^{-4}$ & 0.087 & 0.21 \\
\hline
\end{tabular}
\end{ruledtabular}
\end{table}
We see from the Table that the effects of the introduction of
analytic threshold effects (i.e., the effects of the nondecoupling of $m_c$)
in the calculation of $r_{\tau}$ are $\delta r_{\tau} \alt 10^{-3}$. This is
to be compared with the theoretical value $r_{\tau}^{\rm (LO)} \approx 0.123$\
in the anQCD model, and the full value  $r_{\tau}=0.203$ in the model,
and with the experimental uncertainties $\delta r_{\tau} = \pm 0.004$
[Eq.~(\ref{rtauexp})]. We conclude that, although these effects are
appreciable, they are somewhat lower than the present experimental
uncertainty of $r_{\tau}$.

\section{Summary}
\label{sec:summ}

In this work we presented an analytic QCD (anQCD) model which,
at high squared momenta $|Q^2|$, 
becomes to a high degree indistinguishable from
perturbative QCD (pQCD); nonetheless, at low $|Q^2|$ the 
spacelike couplings $\A_n(Q^2)$ of the model 
mirror correctly the analytic properties of the (to-be-evaluated)
spacelike observables ${\cal D}(Q^2)$ as dictated by the general principles of
the local quantum field theories. The model reproduces
correctly the experimental value of the $\tau$ lepton (nonstrange, $V+A$) 
semihadronic decay ratio $r_{\tau}$, i.e., the only 
low-momentum QCD observable that is well measured and whose
higher-twist effects are small and appear to be under control.
The difference between the analytic $\A_1(Q^2)$ and its perturbative
counterpart $a_{\rm pt}(Q^2)$ [$\equiv \alpha_s(Q^2)/\pi$] is formally
$\sim (\Lambda^2/Q^2)^5$ at $|Q^2| > \Lambda^2$, where
$\Lambda^2 \sim 0.1 \ {\rm GeV}^2$ is the QCD (or: light meson) scale.
The starting point was the construction of the discontinuity
function $\rho_1(\sigma)$ for the analytic coupling $\A_1(Q^2)$.
This  $\rho_1(\sigma)$ is assumed to coincide with its pQCD counterpart
for $\sigma$  above a pQCD-onset scale $M_0^2 \sim 1 \ {\rm GeV}^2$.
On the other hand, the unknown behavior of $\rho_1(\sigma)$ in the
low-$\sigma$ regime is parametrized as a linear combination of two
delta functions.  The underlying pQCD scheme parameter $c_2$ is adjusted 
so that the pQCD-onset scale $M_0$ is either 1.0, 1.25, or 1.50 GeV.
We believe that these three variants of the two-delta anQCD model
represent almost the same physics, since they reproduce the
same value of the key low-momentum observable $r_{\tau}(V+A)$ and,
at the same time, they differ in the value of the 
underlying pQCD (i.e., high-momentum)
renormalization scheme parameter $c_2$, the latter being the only
free input parameter of the model. 
If the value of $c_2$ is further adjusted so that $M_0$ is below
1.0 GeV, the analyticity gets lost at $M_0 \approx 0.9$ GeV.
The model is an extension of the one-delta anQCD model 
of \cite{CCEM} where $c_2=0$ was used.

The main motivation behind the construction of the considered anQCD model
is that with it we can eventually evaluate low-momentum
QCD quantities whose higher-twist contributions are appreciable.
An example are the separate 
vector ($V$) and axial vector ($A$) channel of $r_{\tau}$.
Such evaluations would involve the considered anQCD model together
with the Operator Product Expansion (OPE). 
This application (anQCD+OPE) is consistent, 
due to the very suppressed difference
$\A_1(Q^2) - a_{\rm pt}(Q^2) \sim (\Lambda^2/Q^2)^5$ in the
ultraviolet regime; the latter implying
that the higher-twist terms in OPE, of dimension $d \leq 8$, 
will still be of infrared origin,
in accordance with the philosophy of the ITEP 
(Institute of Theoretical and Experimental Physics) 
group \cite{Shifman:1978bx}.
  
\begin{acknowledgments}
\noindent
This work was supported in part by FONDECYT Grant No.~1095196 
(C.C., G.C.), Rings Project ACT119 (G.C.), 
and Mecesup2 Grant FSM 0605-D3021 (C.A.). 
This work was influenced by the ideas and suggestions of 
our colleague Olivier Espinosa (1961-2010), 
coauthor of our previous work Ref.~\cite{CCEM}.
\end{acknowledgments}

\end{document}